\documentclass[a4paper,fleqn,usenatbib]{mnras}

\usepackage{newtxtext,newtxmath,deluxetable}

\usepackage[T1]{fontenc}
\usepackage{ae,aecompl}

\usepackage{graphicx}	
\usepackage{amsmath}	
\usepackage{amssymb}	
\usepackage[english]{babel}
\usepackage{array}
\usepackage{multirow}
\usepackage{threeparttable}

\newcolumntype{M}[1]{>{\centering\arraybackslash}m{#1}}

\title[X-ray spectrum of NGC 1194]{Elucidating the global distribution of reprocessing gas in  NGC 1194}

\author[T.J.Turner et al.]{
T. J. Turner,$^{1}$\thanks{E-mail: turnertjane@gmail.com}
J. N. Reeves,$^{2,4}$
V. Braito,$^{2,4}$
T. Yaqoob, $^{3,5}$
S. B. Kraemer,$^{2}$
P. Severgnini$^{4}$
\\
$^{1}$Eureka Scientific, Inc, 2452 Delmer St. Suite 100, Oakland, CA 94602, USA\\
$^2$ Department of Physics, Institute for Astrophysics and Computational Sciences, The Catholic University of America, Washington, DC 20064, USA \\
$^{3}$Center for Space Science and Technology, University of Maryland Baltimore County, 1000 Hilltop Circle, Baltimore, MD 21250, USA\\
$^{4}$INAF - Osservatorio Astronomico di Brera, Via Bianchi 46 I-23807 Merate (LC), Italy\\
$^{5}$ NASA/Goddard Spaceflight Center, Mail Code 662, Greenbelt, MD 20771, USA \\
 }

\date{Accepted 2020 Aug 3. Received 2020 July 14; in original form 2020 June 8}

\pubyear{2020}

\begin{document}
\label{firstpage}
\pagerange{\pageref{firstpage}--\pageref{lastpage}}
\maketitle

\begin{abstract}

A  joint {\it XMM-Newton} and  {\it NuSTAR} observation was conducted for the bright, local Seyfert 1.9 galaxy, NGC 1194. 
The hard spectral form of this AGN was modeled using the toroidal reprocessor {\sc mytorus}. The 
decoupled  model form provides a good description of the  spectrum, with reflection arising from gas 
with a global average column density $> 4 \times 10^{24}{\rm cm^{-2}}$ and transmission of the continuum through an 
 order-of-magnitude lower column. 
In this model, the reflection strength is a factor $\sim 3$ higher than expected from a simple torus.  Such a result may indicate that 
much of the intrinsic X-ray continuum is  hidden from view.  An alternative model is that of a patchy torus, where  85\% of sight-lines  are 
obscured by Compton-thick gas and the remaining 15\% by Compton-thin gas.  The patchy torus model is based on a solar abundance of Fe  and is  
consistent with X-ray partial-covering results found in other AGN. That a patchy torus model would 
 relieve the issue with the strength of the reflection signature  is not an intuitive result: such an insight regarding the geometry of the global reprocessing gas 
could not have been obtained using ad hoc model components to describe the spectral form.

\end{abstract}




\section{Introduction}

Mapping the geometry, mass and kinematics of circumnuclear material surrounding supermassive black holes in active galactic nuclei (AGN) is critical to understanding the distribution of observed properties in the local population,  and, more fundamentally, the way in which material and energy are exchanged between the nuclear black hole and host galaxy. 

Recent modeling of a  BAT-selected  {\it Suzaku} sample  \citep{tatum13a} of type 1- 1.9 radio quiet AGN established that, 
contrary to expectations, such sources generally have 
very hard X-ray spectral forms at high energies. 
Even  a  model where simple disk reflection is the only component in the 
spectrum is not adequate to account for {\it NuSTAR} luminosities measured above 10 keV; such models over-predict the Fe line equivalent widths relative to the 
hardness ratios observed \citep{tatum13a}. The most direct way to reconcile the distribution of local  AGN properties is to allow for  a significant component of absorption by Compton thick gas, partially covering the X-ray source, to different degrees for the type 1 - 1.9 cases. This can be reconciled with the type 1 nature of some  AGN  by invoking holes in the gas distribution allowing a time-variable direct view of the nucleus.  
The high column density of gas required in the Tatum et al. study forms a natural extension of  the multi-zoned X-ray absorber, whose signatures are clear in the 0.5-10 keV band \citep[e.g.][]{kaspi02a,reeves04a}. 
Kinematic measurements show that the absorber is predominantly seen to be outflowing \citep[e.g.][]{blustin05a}, tracing some fraction of the accreting 
material that is being blown back into the host, driven by radiation pressure, thermal pressure, magnetohydrodynamic forces or a combination of these. 
X-ray results to date are consistent with   AGN having a cloud filling factor that increases from the pole  view down to the plane of the accretion disk, such that systems viewed close to the plane of the disk have a relatively high probability of there being a large number of clouds in the sight-line 
\citep{nenkova08b}.  The gas distribution required is consistent with clumpy torus models \citep{elitzur06a}. 

While the line-of-sight ionized absorption has been usefully constrained by measurement of the depths of spectral features, the establishment of the global covering factor for the gas,  and the radial location (and thus the total mass)  have proved more challenging. The strengths of observed emission lines can be used in this regard, and in particular, that of the scattered X-ray emission predicted from the reprocessing gas. 

The column density corresponding to an X-ray scattering optical depth of unity is commonly used to set the convention of terminology. This relates to the inverse of the Thomson scattering cross section  ($\sigma_{\mathrm T}$) such that 
${\mathrm{N_H}} \simeq 1/1.2\sigma_{\mathrm T} \simeq 1.25 \times
10^{24} {\mathrm{cm}^{-2}}$ - values above this column density being denoted ``Compton thick'' and those below  it ``Compton thin''. One limitation of work to date has been the  widely-used assumption of an infinite column density for the reprocessing gas, the so called ``Compton-thick'' case. 

Scattering from Compton-thin gas produces strong fluorescent line emission, most notably from the K shell of Fe, and a rich variety of possible spectra (that depend on geometry, viewing angle and column density). When the reprocessor has an infinite column density the  Compton hump  peaks at an  energy that depends on structure orientation and geometry \citep[e.g. see][]{george91a}. 
Physically-motivated models that treat the scattered continuum and Fe K$\alpha$ fluorescent emission-line self-consistently for finite column density material, have been recently released in a form suitable for application to X-ray spectral data  
\citep[e.g.][]{ikeda09, murphy09a, brightman11a, liu14a}. The high energy bandpass of {\it Suzaku} \citep[e.g.][]{yaqoob15a,yaqoob12a} and, later, 
{\it NuSTAR} has allowed these toroidal models to be applied  to Compton-thick AGN, allowing important insight into the 
reprocessor details  \citep[e.g.][]{marchesi19a, brightman15a, balokovic18a}, for which scant information were available previously. 
 
The improved methodologies inherent in the {\sc mytorus}  model allows one to probe the three-dimensional structure of the X-ray reprocessor in a meaningful way. Initial results offer promise for significant progress. For example, application of the {\sc mytorus} model to the type 2 AGN Mrk~3, produces a solution in which the  line-of-sight column density is higher than the global average  column  \citep{yaqoob15a}. 

Models for the local AGN population provide a recent-epoch data point in the fundamental problem of accounting for the 
 Cosmic X-ray Background (CXB). This line of investigation gives us a more realistic estimate of the actual contribution of all 
 AGN (type 1, intermediates and type 2) to the integrated absorbed-light in the CXB and 
 from that one might draw model possibilities for sources seen at higher redshift.  Once this is 
properly accounted for in AGN population synthesis models, we might not need to find a new population of hitherto undetected 
Compton-thick AGN to explain the X-ray background. 

The hardest AGN offer the most insight  into understanding the circumnuclear gas, and 
one of the hardest sources in the  \citet{tatum13a}  sample is NGC 1194.   
This nearby (z=0.0136) Seyfert 1.9 galaxy hosts a water maser, which has allowed a precise measurement of the mass of the black hole at 
$M=6.5\pm 0.3 \times 10^7 M_{\odot}$ \citep{kuo11a}. NGC1194 is also a bright X-ray source, detected in the BAT catalogue. 
 NGC 1194 has been classified as X-ray Compton-thick and has been discussed by 
various authors in the literature, \citep[e.g.][and references therein]{dellaceca08a, georgantopoulos19a, marchesi19a}. Further to this, the source has 
been classified as Compton-thick on the basis of the X-ray to infra-red ratio versus the X-ray hardness ratio  \citep{severgnini12a}. 
Despite the source brightness and interesting properties, NGC 1194 has been poorly studied in the X-ray regime. The 
source has had  only a 16 ks snapshot by {\it XMM-Newton}, 40 ks with {\it Suzaku} \citep{tanimoto18a,tatum13a}
and a 32 ks {\it NuSTAR}  exposure available in X-ray archives. Thus we proposed a joint 
 {\it XMM-Newton} and {\it NuSTAR} observation of the target, and those data form the subject of this paper. 

Section 2 outlines the details of the new observations. Section 3 examines the time variability, along  with a basic spectral parameterization. Section 4 describes the application of a full-covering spherical reprocessor while Section 5 presents detailed modeling using several variations of  the {\sc mytorus} model. Section 6 discusses the results in the context of other information for NGC 1194. 

\begin{table*}
\centering
\caption{Observation Log}
\begin{tabular}{| l | l | l | l | }
\hline

Date & 
Observation ID  & 
Total & 
Flux \\
 & 
 & 
Exposure & 
2-10\,keV  \\
\hline
\hline
{\it XMM-Newton}  &   & & \\
\hline
{\bf 2020-01-16}  & {\bf 0852200101}  & {\bf 62(MOS), 52(pn)}  & {\bf 0.10}  \\
2006-02-19   & 0307000701   & 16   & $ 0.11^1$ \\
\hline
{\it NuSTAR} &   &  & 10 - 70 keV \\
\hline
{\bf 2020-01-17}   &  {\bf 60501011002} &  {\bf 56 (per FPM)} & {\bf 2.18} \\
2015-02-28   &  60061035002 &  32  & $1.80^2$  \\
\hline
\end{tabular} \\
\vspace{-0.7cm}
\tablecomments{Observed fluxes are given in units 10$^{-11}$ erg\,cm$^{-2}{\rm s^{-1}}$ based on the {\it XMM-Newton} pn  and the mean FPM data. Bold face type denotes the sequences presented in this paper. \\
$^1$ \citet{marchesi19a} \\
$^2$ Extrapolated from \citet{marchesi19a}}

\label{tab:data}
\end{table*}%

\section{Observations}

\subsection{XMM-Newton}

{\it XMM-Newton} conducted an observation of  NGC 1194 during 2020 Jan 16  as part of a co-ordinated 
{\it XMM-Newton}/{\it NuSTAR} exposure.   Table~1 gives a summary of the observations. 
The {\it XMM-Newton} observations  were performed in Full Window mode, with the medium filter in place. All data were processed using  {\sc sas} v18.0.0 and {\sc heasoft} v6.23 software.  The  EPIC pn  and MOS source  spectra  were extracted using a
circular region  with a   radius of $35''$  and background spectra  using two circular regions  each with a radius of $35''$.  After cleaning using standard criteria,  {\it i.e.} PATTERN $\leq 12$  for MOS,  $\leq 4$ for pn, pulse invariant channels between 200 and 15000, and FLAG $= 0$, each observation was  filtered  to remove a brief period of high background at the end of the exposure. The resulting  {\it XMM-Newton} event files yielded exposures of $\sim 62$ ks per MOS CCD and $\sim 52$ ks for the pn. 

 The count rates over the 0.5-10.0 keV band were 0.02 ct s$^{-1}$ MOS$^{-1}$ and 0.10 ct s$^{-1}$ for the pn. The background was 7\% of the source count rate in that bandpass.
 We generated the response matrices and  the ancillary response files at the source position   using the SAS tasks \textit{arfgen} and \textit{rmfgen} and the latest calibration available. The EPIC spectra were  binned to a minimum of 50 counts per energy bin: the binned data maintained sampling finer than the spectral resolution of the EPIC CCDs. 
 The {\it XMM-Newton} RGS1 and RGS2 data were reduced  with  the standard {\sc sas} task {\sc rgsproc}, Unfortunately, the source was too faint  in the soft band for a useful RGS spectrum to be accumulated.

\subsection{NuSTAR}

{\it NuSTAR}  carries  two co-aligned telescopes containing Focal Plane Modules A and B (FPMA, FPMB; \citealt{harrison13a} ) covering a useful bandpass of $\sim 3-80$ keV for AGN. 
{\it NuSTAR} observed NGC 1194 on 2020 Jan 17, overlapping  the  {\it XMM-Newton}  exposure (Table~1). 
Event files were created through the {\sc nupipeline} task v0.4.6,  
cleaned  applying  the standard screening criteria, where  passages through the SAA  were filtered out, setting the mode to  ``optimised" 
 in \textsc{nucalsaa}. This  yielded  a net exposure  of $\sim 111$ ks in the summed data from both focal plane modules.   For each of the Focal Plane Modules, source spectra were extracted  from a circular region with a radius of $70''$,  while background spectra were extracted  from a circular region with a $75''$ radius located on the same detector.  
{\it NuSTAR} source spectra were  binned to  100 counts per spectral channel, maintaining a sampling that is 
finer than the spectral resolution of the instruments. 
During 2020 Jan the summed FP modules  yielded a source count rate of  $0.067\pm 0.008$ ct s$^{-1}$  over 3-50 keV. The background level was $\sim 9\%$ of the total count rate.  These rates correspond to observed fluxes  $\sim 1.0 \times 10^{-12} {\rm erg\, cm^{-2} s^{-1}}$  and 
$\sim 1.6  \times 10^{-11} {\rm erg\, cm^{-2} s^{-1}}$  in the 2-10 keV band and 10 - 50 keV band, respectively.

\subsection{General Considerations}

Spectra are analyzed with XSPEC v12.9.1m.  We used data over 0.6 - 10 keV for the pn and MOS and 3 - 50 keV for {\it NuSTAR}. This data range was chosen as our main application here is the  {\sc mytorus}  model, which is not calculated below that energy range. 
All models included a full covering column of gas at a redshift of zero, that covered all components and whose lower limit was set to the  Galactic line-of-sight absorption,
N$_{\rm H,Gal} = 5.53 \times 10^{20}{\rm cm}^{-2}$ \citep{bekhti16a}. This absorber was parameterized using {\sc tbabs} \citep{wilms2000a}.  All other model
components were adjusted to be at the redshift of the host galaxy. 
We assume  $H_o =70\, {\rm km\, s^{-1}\, Mpc^{-1}}$ throughout.

 For the
ionized emitter  we generated model tables using version 2.41 of the {\sc xstar} code
\citep{kallman01a,kallman04a}, allowing abundances that were variable such that they could  float in the fit. {\sc xstar} 
models the reprocessing gas as thin
slabs, with parameters of atomic column density and ionization parameter $\xi$,
defined as $$\xi=\frac{L_{\rm ion}}{n_e r^2}$$ that has units erg\,cm\,s$^{-1}$
and where $L_{\rm ion}$ is the ionizing luminosity between 1 and
1000 Rydbergs, $n_e$ is the gas density in cm$^{-3}$ and $r$ is the
radial distance (cm) of the absorbing gas from the central continuum
source. The spectral energy distribution illuminating this gas was taken to be a simple
power law with $\Gamma=2$. The turbulent
velocity was taken as $\sigma=300$ km s$^{-1}$. 

Parameters are quoted in the rest-frame of the source and
errors are at the 90\% confidence level for one interesting parameter
($\Delta \chi^2 = 2.706$).  In the plots the data are binned more coarsely than allowed in the fit, for visual clarity.  

There is a small  calibration offset  between the different  detectors used 
and so a constant component was allowed in all models, constrained to a range of 0.9-1.1 for the cross-normalization constant between the pn, MOS and the summed {\it NuSTAR} detectors \citep{madsen17a}.

\begin{figure}
\includegraphics[scale=0.43,width=8cm, height=8cm,angle=0]{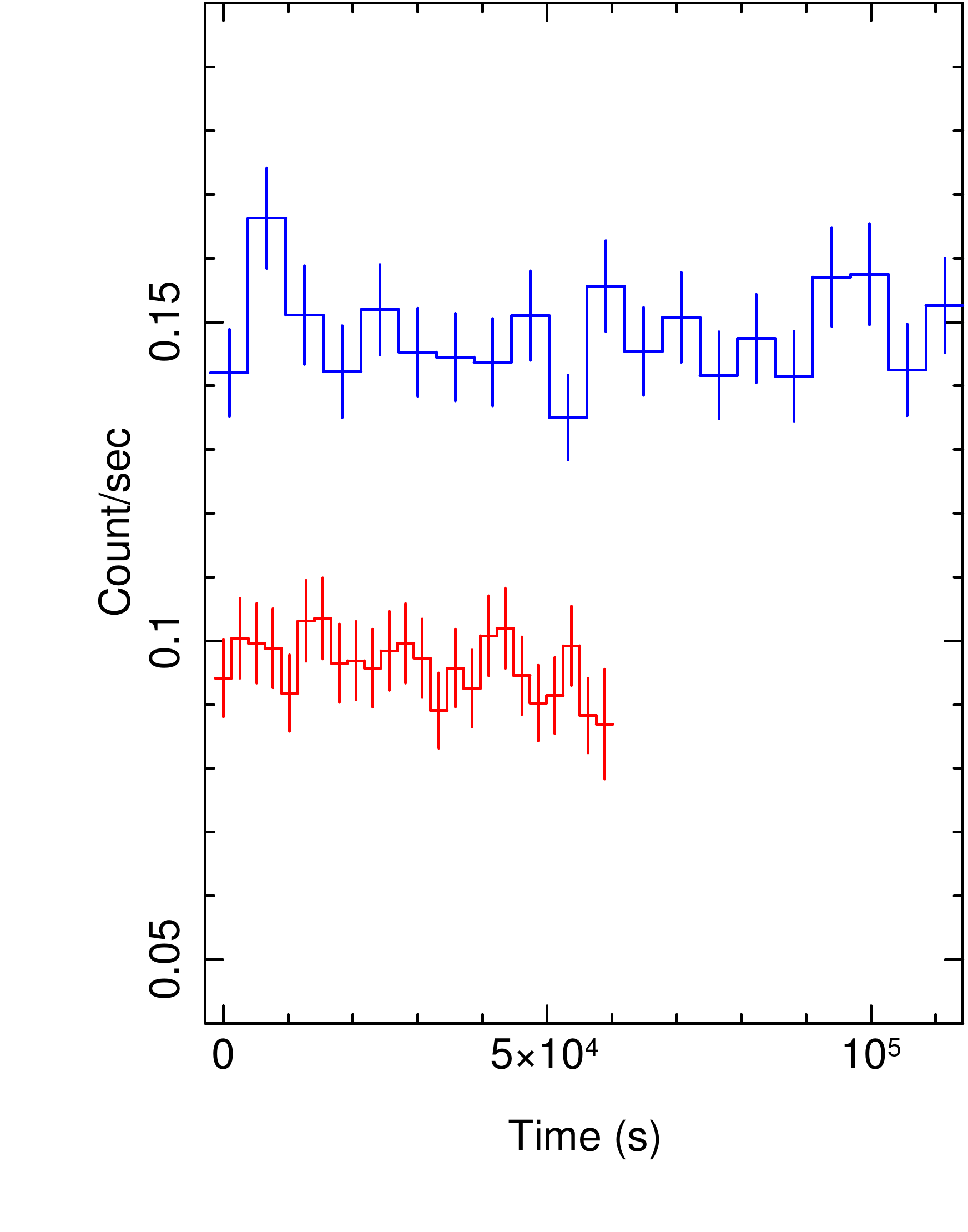}
\caption{The light curves for the {\it XMM-Newton}  pn data (red), over 0.5-10 keV and sampled at 2560 s and the overlapping {\it NuSTAR} data (blue), covering 3-70 keV binned at the  orbit timescale (5814 s). The X-axis shows the time from the start of the XMM-Newton observation.}

\label{fig:pn_rg}
\end{figure}

\begin{figure}
\includegraphics[scale=0.43,width=8cm, height=8cm,angle=0]{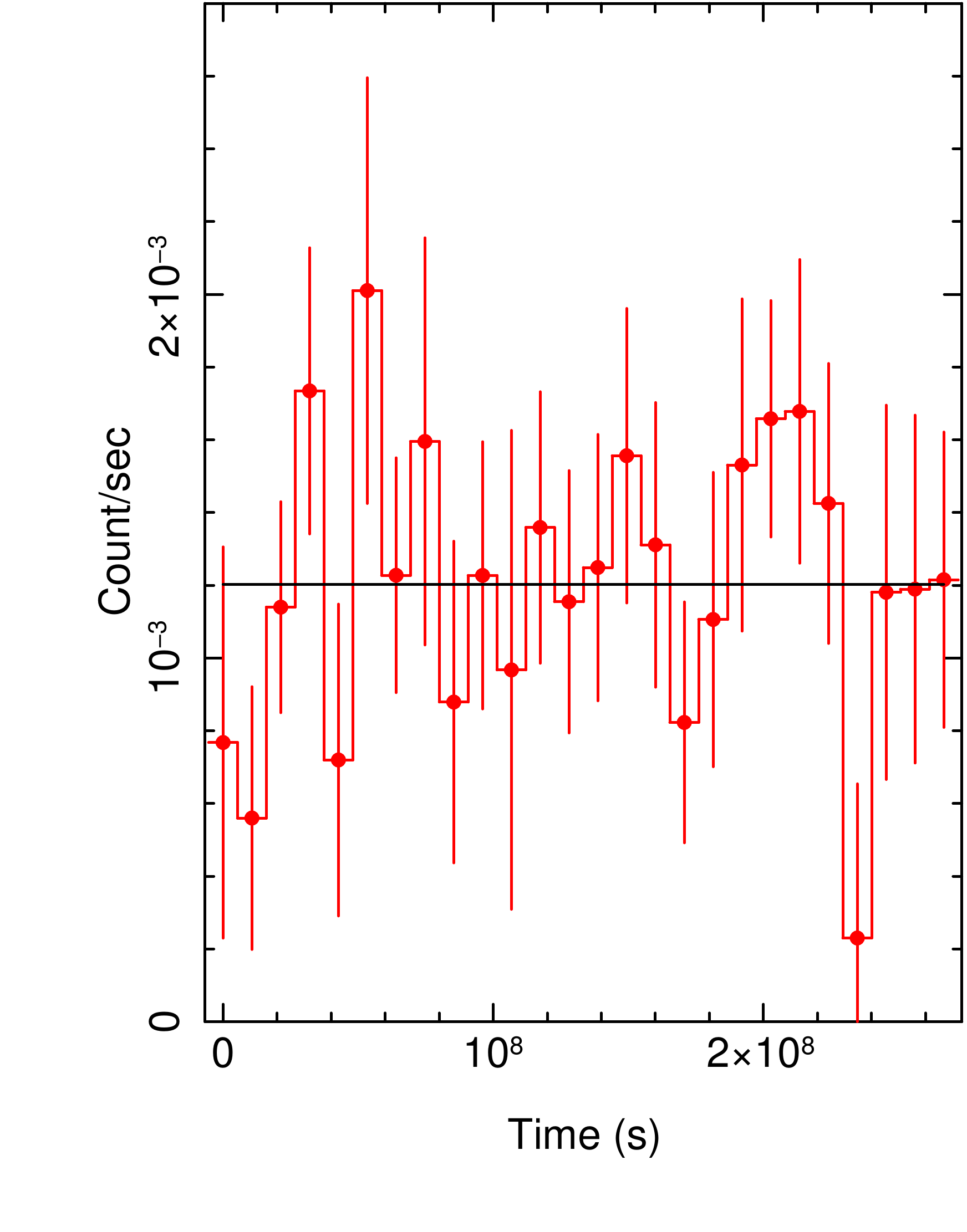}
\caption{The  105 month BAT light curve sampled at 4 month intervals. The zero-time point corresponds to 2004 Dec. The horizontal line represents the best fitting constant value.}
\label{fig:BAT_curve}
\end{figure}

\begin{figure}
\includegraphics[scale=0.43,width=6cm, height=8cm,angle=-90]{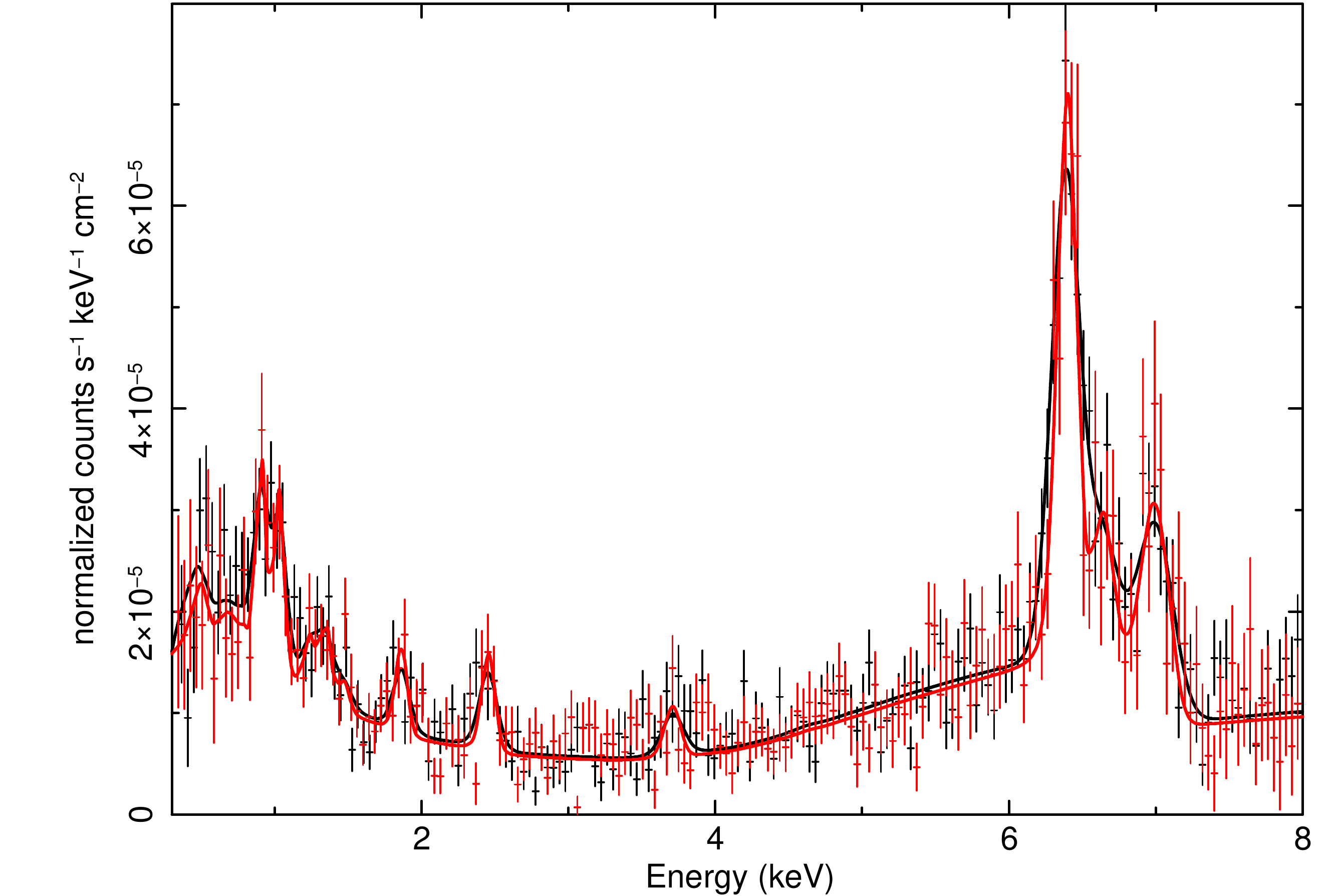}
\caption{Emission lines in the EPIC data from the pn (black) and MOS (red). The model is represented by a solid line (see text for details).}
\label{fig:lines}
\end{figure}

\begin{figure}
\includegraphics[scale=0.43,width=6cm, height=8cm,angle=-90]{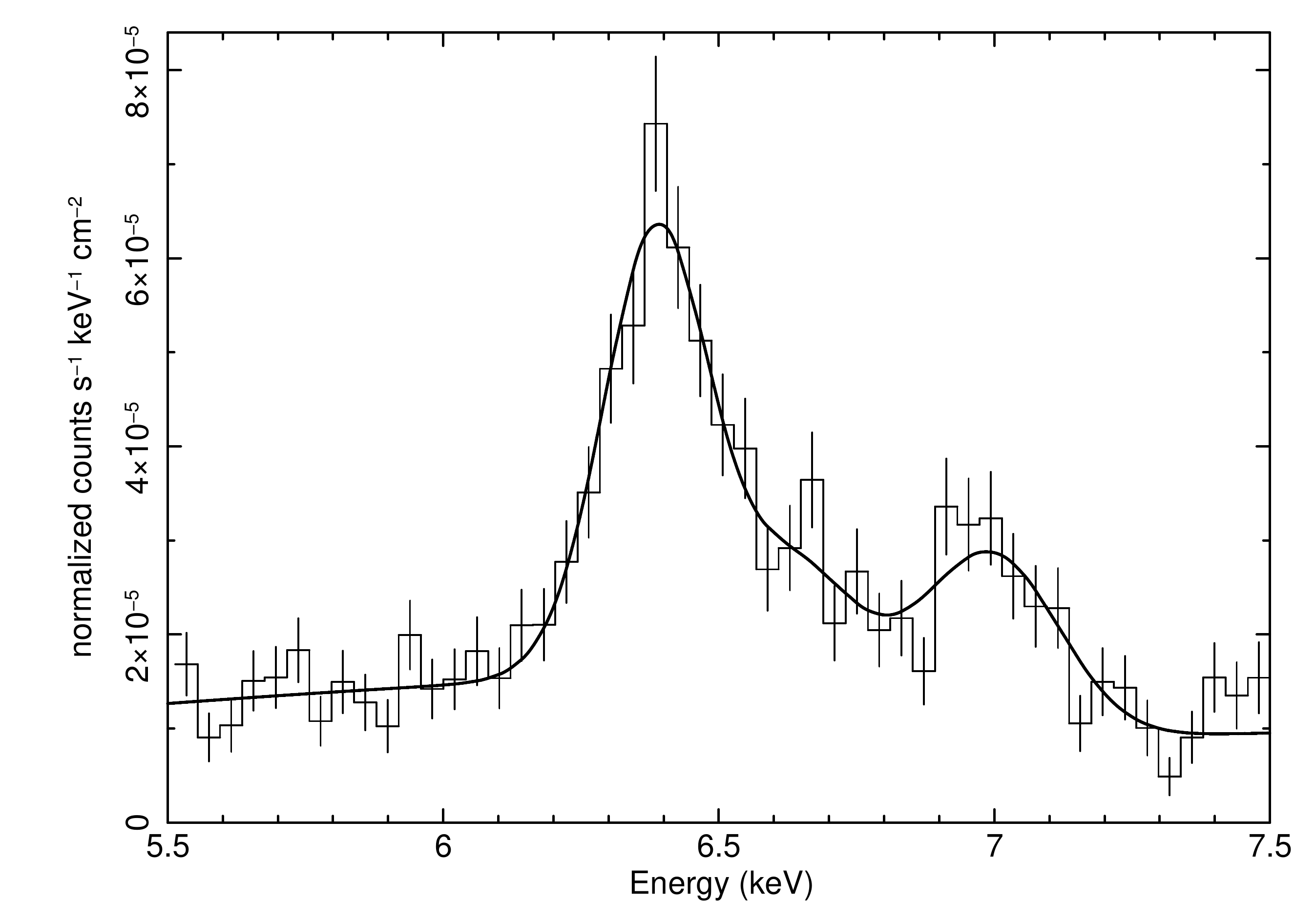}
\includegraphics[scale=0.43,width=6cm, height=8cm,angle=-90]{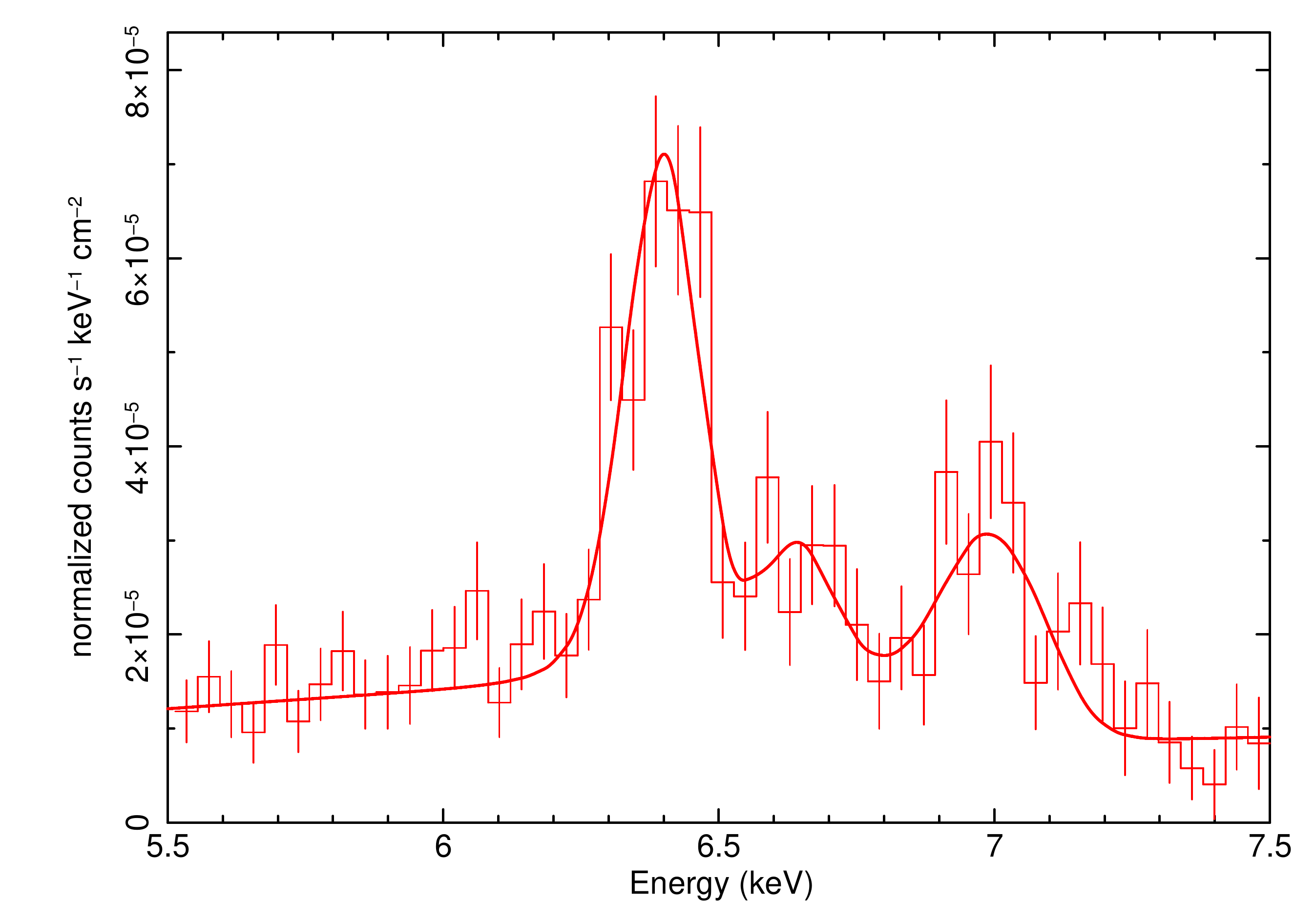}
\caption{The counts spectrum in the Fe K band, for the pn (top, black) and MOS (bottom, red) data. The model line is shown as for Figure~3} 
\label{fig:fek}
\end{figure}

\begin{figure}
\includegraphics[scale=0.43,width=8cm, height=8cm,angle=-90]{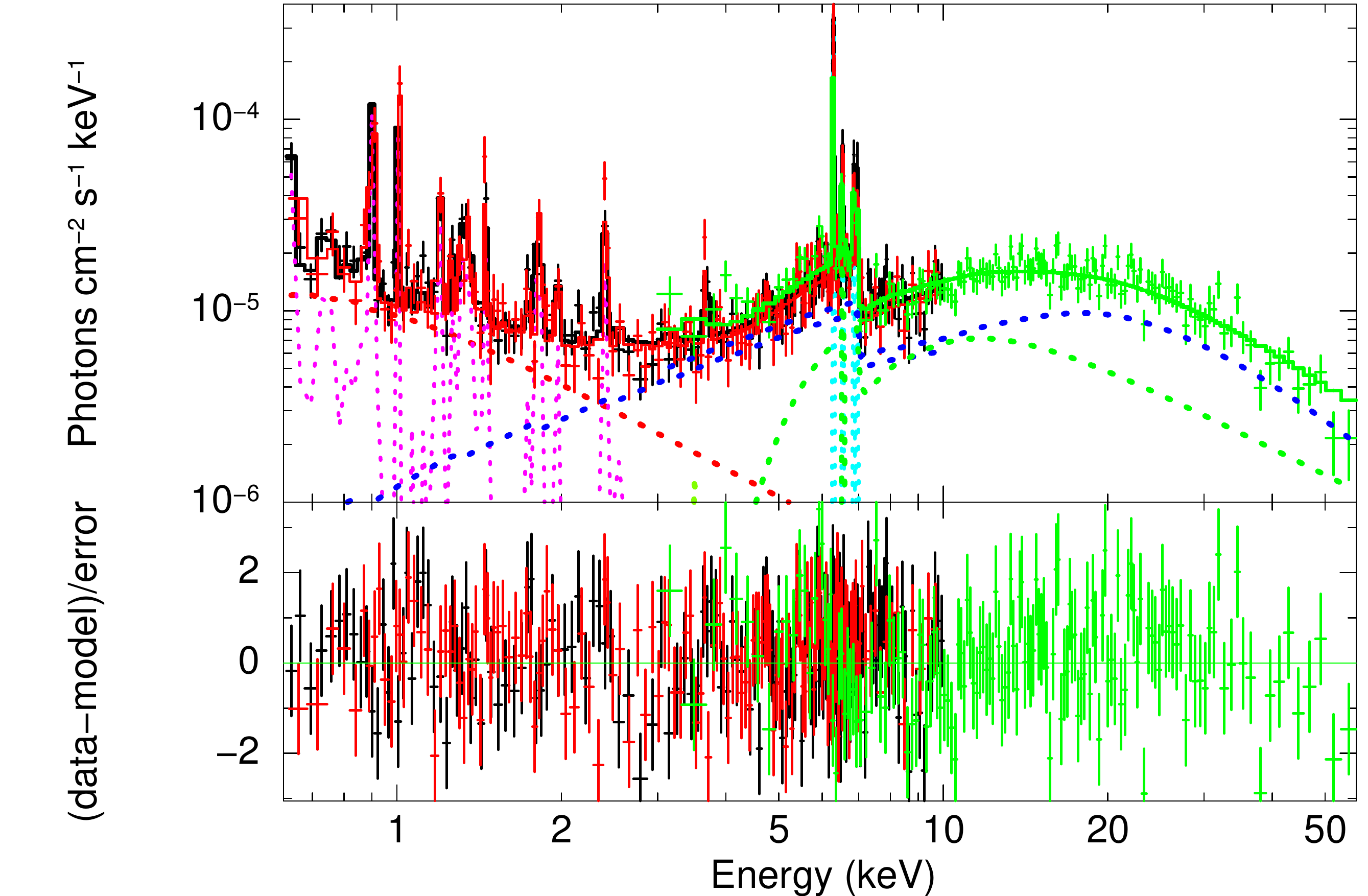}
\caption{The 2020 Jan broad-band spectrum of NGC 1194, fitting the XMM-Newton pn (black), MOS (red)  and NuSTAR (green) data together. The model 
components are an absorbed power law (blue), an {\sc xstar} emission line table (magenta), 
{\sc pexrav} for a reflection component (green dotted line), a scattered fraction of the unabsorbed power law 
(red dotted line) and Fe K emission (aqua).}
\label{fig:broadfit}
\end{figure}

\begin{table*}
	\centering
	\caption{X-ray Line Emission}
	\label{tab:table}
	\begin{tabular}{  | llll  |} 
\hline
E & ID & Flux  & EW   \\  
 keV &   & $\frac{  {\rm photons\, cm^{-2} s^{-1} }  }  {10^{-6}   }$ &  eV  \\
\hline
0.91$\pm0.01$ & Ne {\sc ix} & 2.77$\pm0.64$ & 141 \\
1.04$\pm0.01$ & Ne {\sc x} &2.21$\pm0.52$ & 134 \\
1.23$^{+0.04}_{-0.03}$ & Mg {\sc i} K$\alpha$ & 0.81$\pm0.35$ & 61 \\
1.34$\pm0.02$ &  Mg {\sc xi} &1.00$\pm0.34$ & 84 \\
1.47$\pm0.05$ & Mg  {\sc xii} & 0.44$\pm 0.36$ & 40 \\
1.88$^{+0.01}_{-0.03}$ & Si  {\sc xiii} & 1.19$^{+0.39}_{-0.22}$ & 136  \\
2.45$\pm 0.02$ & S {\sc xv} & 1.27$\pm0.41$ & 169 \\
3.72$^{+0.17}_{-0.05}$ & Ca K$\alpha$ & 0.50$\pm0.32$ & 67 \\
6.39$\pm0.01$ & Fe K$\alpha$ & 11.12$^{+0.81}_{-0.92}$  & 550 \\
6.66$\pm0.04$ & Fe {\sc xxv}& 2.28$^{+0.74}_{-0.65}$  & 63 \\
6.96$\pm0.04$ & Fe {\sc xxvi} & 1.60$^{+0.69}_{-0.67}$  & 78 \\
7.05 (fixed) & Fe K$\beta$  &1.50 & 92 \\
\hline
\end{tabular}
\end{table*}

\section{Initial Analysis}

\subsection{Flux Stability}

Light curves were constructed for the source, extracting the counts through the source regions detailed in the previous section. The 
{\it XMM-Newton} pn data were sampled using 2560 s bins and in the full band of 0.5-10.0 keV. The {\it NuSTAR} data  covering 3.0-70.0 keV were used, sampled on the orbit timescale of 5814 s and data from the two FP modules were summed. 
The {\it NuSTAR} exposure was initiated 920 s after the {\it XMM-Newton} observation was started. Figure~1 shows the light curves for the {\it NuSTAR} and  {\it XMM-Newton} sequences. 

Comparison of the hard-band flux from our {\it NuSTAR} observation and the archived data from 2015 indicates there to be no significant variability on timescales of years. To further check the count rate trend on long term timescales, we extracted the light curve from the {\it Swift} Burst Alert Telescope (BAT), based on 105 months of sky survey data and available online at \verb+https://swift.gsfc.nasa.gov/results/bs105mon/+ . 

This BAT light curve, sampled at 4 months, covers 14-195 keV and includes observations carried out between 2004 December and 2013 August. The source flux shows no significant variability over that $\sim 9$ year baseline (Figure~\ref{fig:BAT_curve}).  Fitting the data to a constant 
flux model produces  a $\chi^2_r =  22/25$. 

\subsection{Basic characterization of the X-ray spectrum}

Initially just the XMM-Newton (pn and MOS) data were fitted to parameterize the soft X-ray continuum and the multiple emission lines that were found to be present both at soft X-rays (Figure~\ref{fig:lines}) and at Fe K (Figure 4).  The line energies are consistent with 
blended Ne {\sc ix-x} in the 0.9-1 keV regime, 
Mg {\sc xi} at $\sim 1.3$ keV, Si {\sc xiii} at $\sim 1.8$ keV and  S {\sc xv} at $\sim 2.4$ keV. Further to these, a line identified at $\sim 3.7$ keV is likely from 
Ca K$\alpha$.  In the Fe K band, emission components include contributions from neutral gas at 6.4 keV, He-like and H-like Fe at 6.67 keV and 6.97 keV,  
plus  Fe K$\beta$ emission at 7.06 keV that is modeled as a fixed 13.5\% of the fitted Fe K$\alpha$ line flux (Table 2; Figure~\ref{fig:lines}). The lines are all consistent with being narrow, with $\sigma < 30$ eV for the Fe K$\alpha$ component at 6.4 keV. The observed mix of ionized and neutral line components are not consistent with an origin in a single gas zone. The lines were fitted using Gaussian profiles, and the line parameters are summarized in Table~2. The baseline model over which the lines were fit, had the following form:

$$N_{H,Gal}  \times  [   N_H  \times [(1 - f) \times po] + f  \times [po  + zg(12) ]     ] $$

where  $N_{H,Gal}$ is the Galactic column, $po$ is the power law (assumed to fold over at 500 keV). $f$ is the covering fraction of the absorbing gas, $N_H$ is an intrinsic  absorber that covers only a fraction (1-$f$) of the continuum. 
The intrinsic absorber is represented by the {\sc xspec} model {\sc phabs}, along with the Compton scattering correction for that column density, applied using 
{\sc cabs}.  zg(12)  represents the twelve individual Gaussian line components. This model fit yielded 
 $f=0.078\pm0.008$, $\Gamma=1.08\pm0.17$, $N_{H}= 3.89^{+0.46}_{-0.45}  \times 10^{23} {\rm cm^{-2}}$. This model provided a good fit to the broad X-ray form of the {\it XMM-Newton} data, with $\chi^2=316/297\, d.o.f$. 
 
The absence of O lines in the spectrum is surprising, and so we investigated this further, fitting the pn and MOS spectra  jointly, using an emission table generated from {\sc xstar}, allowing variable abundances. This table was assumed to be absorbed by the Galactic column.  The host galaxy of NGC 1194 is viewed at high inclination, $\frac{b}{a}=0.45$ \citep{veroncetty06a}.
In order
to constrain the column density of gas within the host galaxy, comprising the interstellar medium (ISM) of  the host,  we retrieved the fluxes
of H$\alpha$ and H$\beta$ from SDSS (Ahumada et al. in prep.). The observed
ratio of H$\alpha$/H$\beta$ was $\approx$ 17.5. Assuming an intrinsic ratio of 2.9 
\citep{osterbrock06a} 
and the Galactic extinction curve 
\citep{savage79a} 
we calculated
a reddening of $E(B-V) = 1.6$, which would arise in an H~{\sc i} column density of $\sim 8.5 \times 10^{21}$ cm$^{-2}$
\citep{shull85a}. 
Since it is likely that some fraction of the gas in the host disk is ionised, 
we assumed an ISM column density of 10$^{22}$ cm$^{-2}$.

The soft-band lines in NGC 1194 have recently been resolved and found  to arise in an extended X-ray region of gas (Braito et al. in prep.), therefore we placed the host galaxy absorption in front of the emission table in the model. However, the strength of the soft flux in this source, and the shape of the soft-band spectrum, are not consistent with the entire nuclear spectrum being obscured by an  ISM column density as high as  $N_{H,ISM} \sim 10^{22}$ cm$^{-2}$. This raises the question as to what elements of the observed spectrum are obscured by this ISM gas component. In a similar AGN,  NGC~7582, \citep{braito17a} find the dust lane to obscure part of the spatially extended narrow line region gas, but not the nuclear emission.  In NGC~7582 the dust lane is responsible for the suppression of the emission lines below 1 keV, which arise from the AGN NLR. 

The detailed morphology of NGC 1194 is under study (Braito et al. in prep). It appears that, similar to NGC~7582, the ISM gas lane may obscure part of the NLR and thus the most appropriate application of that absorption in the spectral fit, is to apply it only to the soft-band line-emission gas.  We proceed under this assumption in our spectral fits. 

The emitter column density was also fixed at $N_H = 1 \times 10^{22} {\rm cm^{-2}}$,  and the fit returned log $\xi=1.77^{+0.08}_{-0.17}$. Abundances were allowed to float and found to be $A_O >0.57$, $A_{Ne}=0.84^{+0.65}_{-0.31}$,  $A_{Mg}=0.12^{+0.37}_{-0.12}$, $A_{Si}=0.70^{+0.66}_{-0.27}$, $A_S=2.76^{+0.45}_{-1.20}$ of solar.  From this fit, it is evident that O line emission is suppressed by the ISM absorber, and once that is taken into account, the data are consistent with solar abundance of O in the emitting gas. Freezing all abundances at 1.0 increased the fit statistic by $\Delta\chi^2=25$, with the increase driven by $A_{Mg}$ and $A_S$, which are not consistent with solar abundance in this fit. This may indicate either a true deviation from solar abundances, or an inadequacy in the ISM modeling. No further progress can be made on this question, with the data in hand.  (This model gas zone cannot account for the Ca and Fe  K-shell lines, which are addressed in later fits.)

Following this simple parameterization, the  {\it NuSTAR} data were then included in a joint fit with the pn and MOS spectra. The baseline model comprised the following form: 

$$N_{H,Gal} \times N_{H1} [N_{H2} [(1 - f) \times po] + f  \times (po + ref + N_{H,ISM}*{\tiny EMIS} + zg(5) )]$$

where   {\sc emis} is the emission profile from the {\sc xstar} table. {\it Ref} is reflection from gas out of the line-of-sight - represented here by the {\sc xspec} model 
{\sc pexrav}, assumed to be viewed orientated at  60$^{\rm o}$, to be illuminated by the fitted power law  and have a cosmic abundance of Fe. 
 Other components are designated as for the previous fit. 
 zg(5)  represents the five individual Gaussian line components not accounted for by the {\sc xstar} table or by {\sc pexrav}  i.e. the  two ionized Fe lines, the neutral Fe K $\alpha$ and K$\beta$ lines and 
Ca K$\alpha$ (Table 2): although these lines might arise in the reflector, they are not included in the {\sc pexrav} model. 
This model fit yielded 
 $f=0.021\pm0.005$, $\Gamma=1.51\pm0.04$, $N_{H,1}= 6.91^{+2.19}_{-6.91}  \times 10^{20} {\rm cm^{-2}}$, $N_{H,2}= 8.74^{+1.36}_{-1.27} \times 10^{23} {\rm cm^{-2}}$. This model provided a good fit to the broad X-ray form (Figure~\ref{fig:broadfit}), with $\chi^2=586/450\, d.o.f$. 
In this fit, the reflection component is relatively strong, yielding R$=1.52^{+0.32}_{-0.47}$ against the continuum. R=1 represents a source that subtends 2$\pi$ steradians to the continuum, such as one might expect for a standard geometrically thin accretion disk. High R values are sometimes explained as being due to a prior episode of  higher continuum with time delays leading to the chance capture of a relatively high reflection signature. That is not likely applicable to NGC 1194, which does not show evidence for continuum variability (Figure~1, Table~1). 
The emission lines are narrow, with an upper limit of $\sigma < 30$ eV: this suggests an origin from the outer broad line region (BLR) or torus. The observed line fluxes and equivalent widths are consistent with those detailed in Table~2.

\section{Spherical Reprocessor Model}

Next we investigated more physically motivated models for the X-ray reprocessor around NGC 1194. First we compared the data to a model utilizing 
reprocessing in a uniform full-covering sphere of gas. The reprocessor was implemented using an additive table named {\sc sphere0708.fits} 
that is a publicly available table originating from the calculations of \citet{brightman11a}. 
 The sphere is assumed to be illuminated by a power law continuum with a high energy cut-off at 500 keV.  
 The Fe abundance was allowed to be free and the other metal abundances (that are controlled by a single scaling parameter) were set to solar, using \citet{anders89a}.  The photoelectric absorption cross-sections  used for the sphere are those of \citet{verner96a}.  In generating the table, \citet{brightman11a}  self-consistently calculate the Fe K$\alpha$ and K$\beta$  emission lines from the gas, and  incorporate  K$\alpha$ emission from other elements, including Mg and Ca. In this model, the Fe line emission, scattered X-rays  and absorption opacity cannot be decoupled. The other emission lines in the data (from ionized gas) are modeled using  eight individual Gaussian components and fitted line parameters are consistent with those tabulated previously (Table~2). Galactic absorption was added to the model, and the final model form was  $$N_{H, Gal} \times [pow +  sphere0708  + zg(8) ]$$
 
\begin{figure}
\includegraphics[scale=0.43,width=8cm, height=8cm,angle=-90]{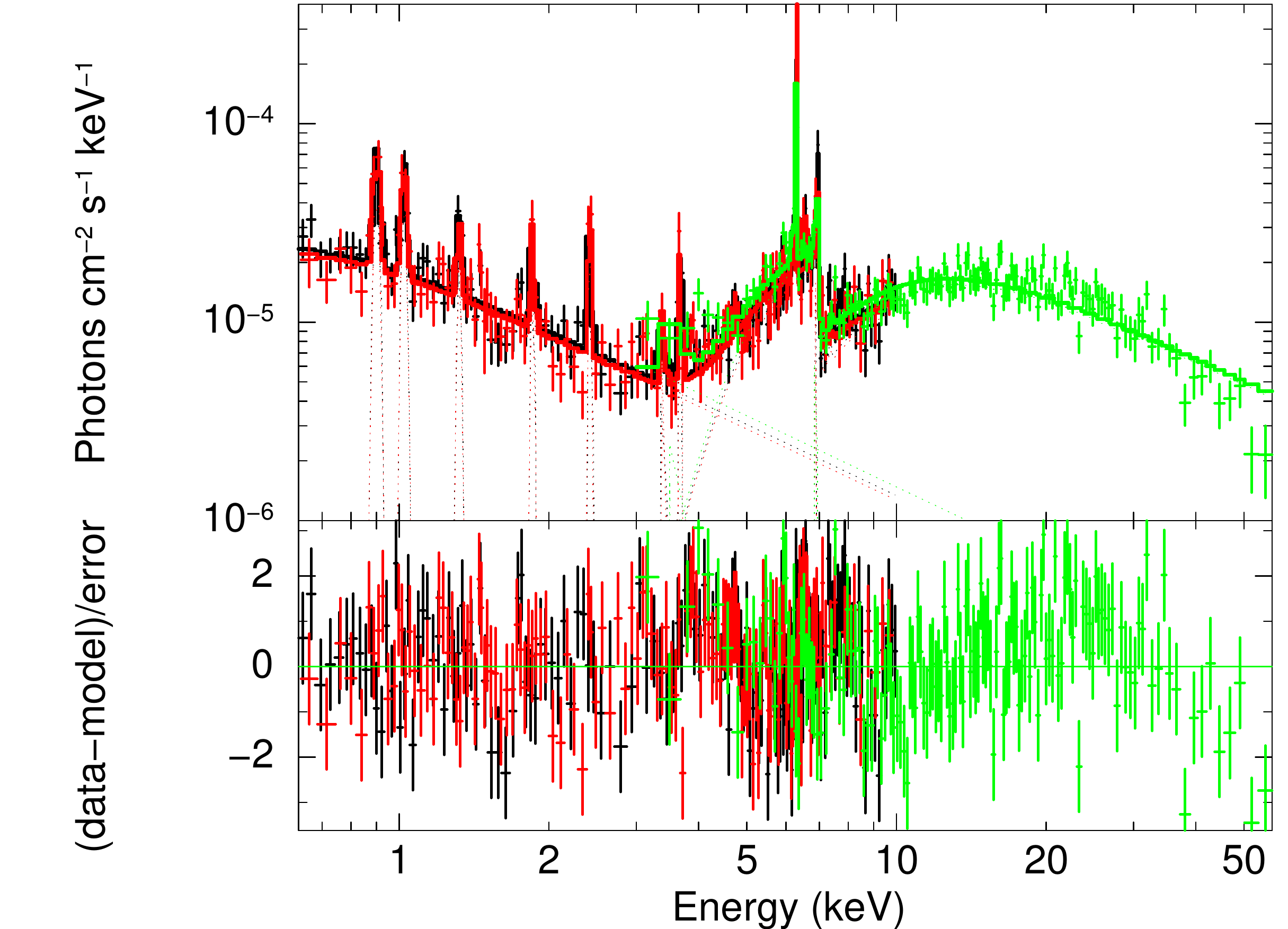}
\caption{The spherical reprocessor of \citet{brightman11a}, compared to the data . Colors as for Figure~\ref{fig:broadfit} }
\label{fig:bn11}
\end{figure}

The model yielded $\chi^2=598/437$ d.o.f. The local column was found to be in excess of the Galactic value, with $N_{H,Gal}=9.07^{+4.66}_{-4.25} \times 10^{20}{\rm cm^{-2}}$, covering a continuum with 
$\Gamma=1.21^{+0.05}_{-0.04}$, with a column density of $N_H = 4.76^{+0.25}_{-0.32} \times 10^{23}{\rm cm^{-2}}$ and Fe abundance 
 $A_{Fe}=1.71^{+0.13}_{-0.12}$.  
 The fit does not provide adequate curvature above 10 keV to explain the shape of the high energy hump (Figure~\ref{fig:bn11}), and also yields an unrealistically flat  power law index for the illuminating continuum.  With these two limitations, we do not consider the spherical reprocessor further. 
 
\section{MYTorus }

The toroidal X-ray reprocessor model, {\sc mytorus}   \citep[][]{murphy09a,yaqoob10a}, is illuminated by a central X-ray continuum source. The global covering factor of the reprocessor is 0.5, corresponding to a solid angle subtended by the structure at the central X-ray source of 2$\pi$  (an opening half-angle of 60$^{\rm o}$). {\sc mytorus} self-consistently calculates the Fe K$\alpha$ and Fe K$\beta$ fluorescent line emission and includes the effects of absorption and Compton scattering on both the  continuum and lines.  {\sc mytorus} offers the ability to leverage constraints on the Fe K line emission and spectral curvature  together, to obtain both the global and line-of-sight column density. The model is  therefore one of the most powerful available for fitting X-ray spectra. 
 Although  the nominal covering factor is fixed at 50\% of the sky as seen by the continuum source, one can model other physical possibilities using the `decoupled' mode that allows for a different line-of-sight column to the gas responsible for reflection (the Fe line emission strength can  still be linked self-consistently to the column of the reflector).  Such a decoupling is often required in spectral analyses of high-quality broadband X-ray spectra, because the column density in and out of  the line-of-sight column density are not always the same \citep[e.g.][]{risaliti09b,marchese12a,yaqoob15a}.
  
In our fits, a single power law was assumed to be seen in transmitted and scattered light, and this was implemented by linking the photon indices of these components in all fits. The general model form was 

$$N_{H} \times [(MYTZ*pow) +  pow + MYTS + gs*MYTL + zg(10) ]$$

where {\it MYTZ} designates the line-of-sight absorption, {\it MYTS} the scattered X-rays and {\it MYTL} the line emission (both from material out of the line-of-sight, hereafter referred to as the `global' gas), from {\sc mytorus}. zg(10) represents ten individual Gaussian lines (Table~2) present in the data that are not accounted for by the {\sc mytorus} model.  {\it gs} is a Gaussian smoothing component convolved with the torus neutral Fe line profile, using a kernel of $\sigma=30$ eV at 6 keV. The power law continuum is split over two sight-lines, one is absorbed my {\it MYTZ} and the other is absorbed only by $N_H$, a full-covering column of gas whose lower limit is set at the Galactic line-of-sight absorption, but whose upper limit is not constrained.

 \subsection{Coupled Model Fits}
 
Taking the ``Coupled Model'' to mean one for which the line-of-sight  and global column densities are tied together, we performed two variations on this fit. First the angle of the  global gas was  fixed at $\theta=90^{\rm o}$ and the normalization of the global-scattered light and line emission were directly tied to the illuminating continuum. This model did not provide an acceptable fit to the data, yielding $\chi^2=709$/438 d.o.f. with 
$N_{H,MYT}=7.2^{+0.2}_{-0.4} \times 10^{23}{\rm cm^{-2}}$, while the full covering gas component gave 
$N_{H} =2.04^{+0.49}_{-0.44}\times 10^{21}{\rm cm^{-2}}$, exceeding the Galactic line-of-sight value.  $2.9$\% of the total intrinsic continuum flux escaped without absorption by the torus (Table~3).

\begin{figure}
\includegraphics[scale=0.43,width=8cm, height=8cm,angle=-90]{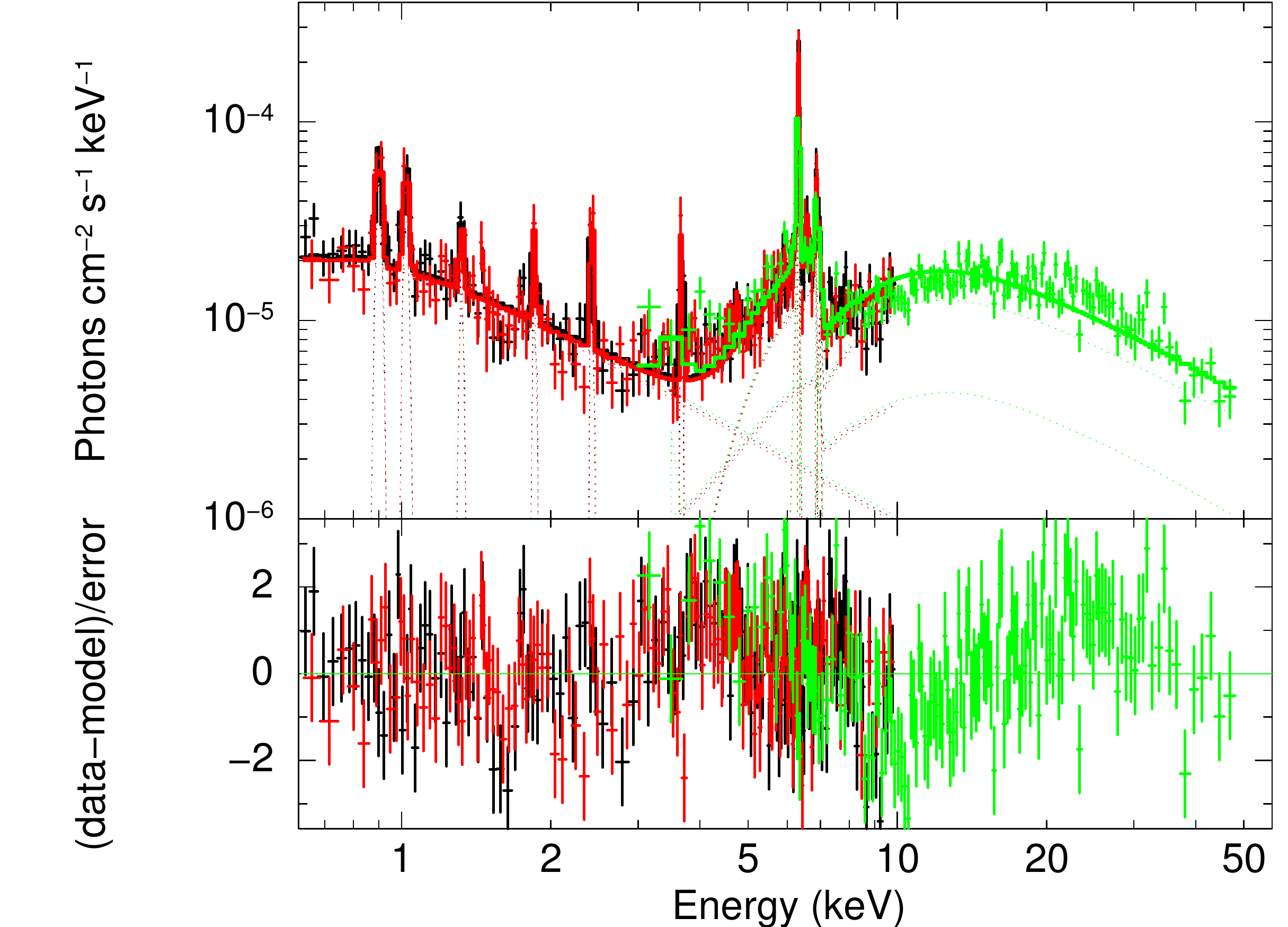}
\caption{The {\it XMM-Newton}/{\it NuSTAR} data from 2020 with fit and residuals for Coupled Model II. Colors as for Figure~\ref{fig:broadfit}.}
\label{fig:CMii}
\end{figure}

A useful variation on the strictly-coupled model is to allow a modest offset of orientation angle for the global gas, and we investigated this possibility, 
setting $\theta=70^{\rm o}$ and again, the normalization of the global-scattered light and lines were tied  directly to the illuminating continuum. This fit yielded $\chi^2=698$/438 d.o.f. In this case the torus column density was 
$N_{H,MYT}=8.4^{+0.4}_{-0.3} \times 10^{23}{\rm cm^{-2}}$ while the full-covering layer had 
 $N_{H} =1.72^{+0.47}_{-0.43} \times 10^{21}{\rm cm^{-2}}$. Again we require a small escape of flux without absorption by the torus, in this case, $\sim 2$\%. While the fit is superior to the previous coupled model, the model has too shallow of a high-energy hump to account for the data (Figure~\ref{fig:CMii}). 

Previous work \citep{georgantopoulos19a, marchesi19a} had found the coupled {\sc mytorus} model to provide a good fit to NGC 1194. \citet{marchesi19a} found a column density that was consistent with that in our coupled {\sc mytorus} fit. To make a direct comparison, we reduced the archived {\it NuSTAR} data from {\sc obsid} 60061035002 taken during 2015 Feb 28 (Table~1). Those data were reduced following the same method as for our new observation. We then applied the coupled {\sc mytorus} model  derived from our 2020 data and fit it to  the 2015 {\it NuSTAR} data. We confirmed the adequacy of the coupled model for this short exposure using {\it NuSTAR} alone  \citep[c.f.][]{georgantopoulos19a, marchesi19a}, this fit yielded $\chi^2=115/92$ d.o.f. We then took the coupled model fit to the archived 2015 data and re-applied it to our new joint {\it XMM} and {\it NuSTAR} data. This yielded $\chi^2=3356/435$ d.o.f. The coupled model fails to account for the data below 3 keV, where the observed flux is much higher than the model prediction. We return to this point in the discussion.

In conclusion, the coupled models provide a statistically unacceptable fit to the combioned {\it XMM-Newton} and {\it NuSTAR} data, and a visual inspection of the fits shows this to be due to the fact that the curvature below 10 keV requires a lower column density than that which is required to explain the observed Fe K emission line and overall reflection signature strength. 

\subsection{Decoupled Model Fits} 

The standard, coupled configuration representing a toroidal reprocessor with an opening angle of 60$^{\rm o}$ fails to account for the relative strength of the hard spectral excess. So-called decoupled models allow for the  components in and out of the line-of-sight to be separated, such that one may view the transmitted continuum through a different column density than that from which the X-rays are scattered. So these column densities  are independent in the fit.
The toroidal global scattered gas -({\it MYTS} $+$ {\it MYTL}) that is out of the line-of-sight, is  fixed at an inclination angle of  $\theta=0^{\rm o}$ and is viewed directly without attenuation.  The direct or transmitted component ({\it MYTZ}) is fixed at 90 degrees inclination and is attenuated by the full column of gas in the {\sc mytorus} model.  

The column densities of the line-of-sight and global gas were decoupled. 
As the strength of the globally-scattered component has been established to be strong in this source, we went directly to a fit that allowed the global gas component to be viewed 
at  $\theta=0^{\rm o}$. This fit gave  $\chi^2=596$/437 d.o.f.  (Table~3 column 4, Decoupled {\sc i}). 
A photon index $\Gamma \sim  2$ was recovered. The global scattering gas has a column density 
$> 7.6 \times 10^{24}{\rm cm^{-2}}$, while  the line-of-sight absorption was $N_{H, MYT} \sim 1.20^{+0.08}_{-0.05} \times 10^{24}{\rm cm^{-2}}$.  
The relative flatness of the model compared to the data around 20 keV suggests that this configuration of model components under-predicts 
the strength of the reflected light component. 

\begin{figure}
\includegraphics[scale=0.43,width=8cm, height=8cm,angle=-90]{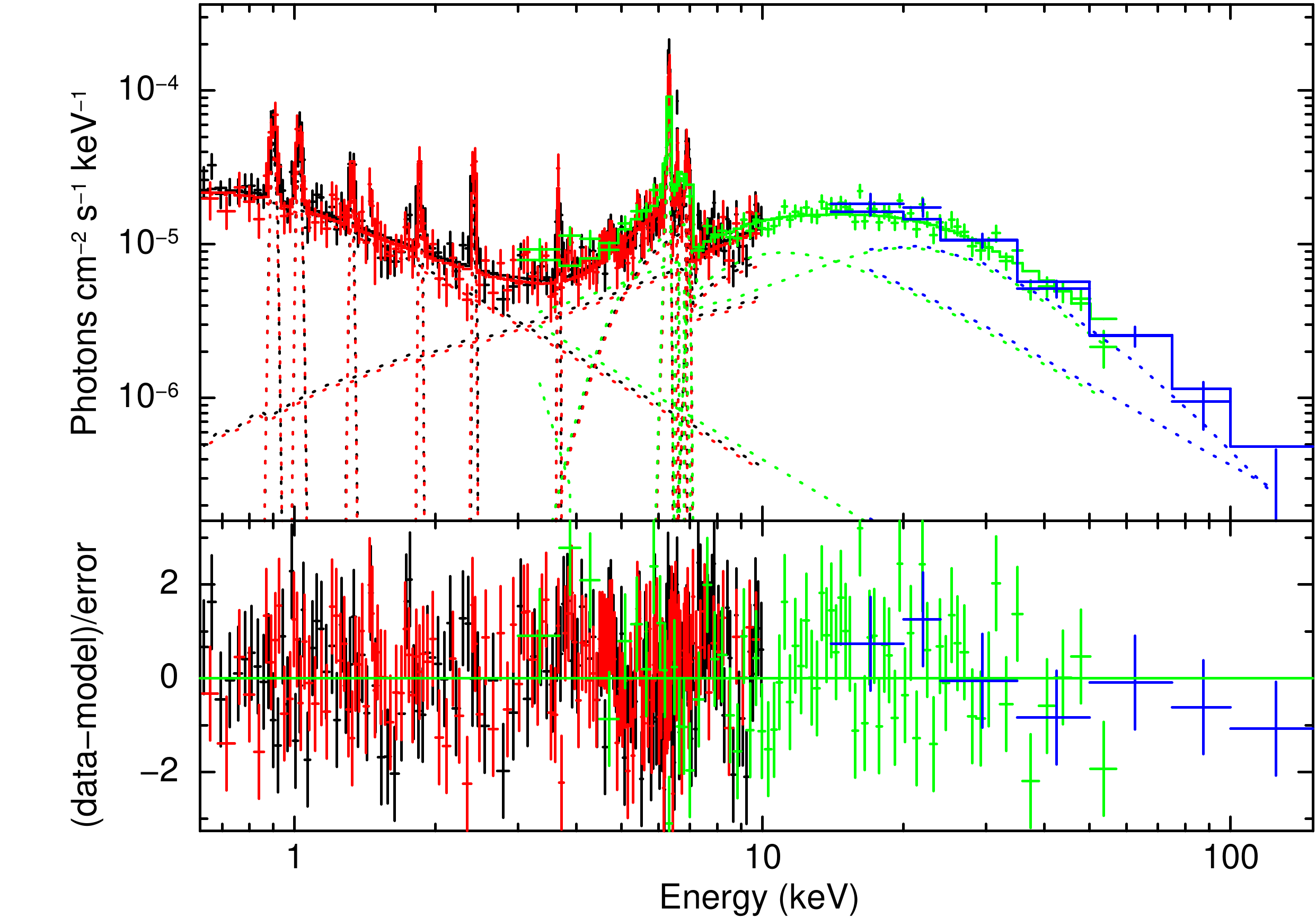}
\caption{The BAT 105 month spectrum (blue points) compared to the {\it XMM-Newton}/{\it NuSTAR} data from 2020, and  Decoupled Model II. Colors as for Figure~\ref{fig:broadfit}. }
\label{fig:BATspec}
\end{figure}

All fits thus far have had the  normalization of the global-scattered light and lines tied  directly to the illuminating continuum.  Once a scaling constant is  introduced for the reflected light and line emission,  one can account for the inadequacy of the previous model. The best fit scalar was a factor of 
$\sim 3$ above the nominal reflection strength (Table~3, column 5, Decoupled {\sc ii}), and this addition to the model yielded a significant improvement to the fit, with  
$\chi^2=520$/436 d.o.f.  (Figure~\ref{fig:BATspec}).  In this case, a photon index $\Gamma \sim  1.9$ was recovered. The global reprocessing gas has a column density 
$> 4 \times 10^{24}{\rm cm^{-2}}$, an order-of-magnitude higher  than the line-of-sight absorption.  

To check the source spectral constancy over long timescales we  overlaid the average 105-month BAT spectrum onto the best-fit {\sc mytorus} decoupled model II. The BAT data are consistent with the current epoch {\it NuSTAR} data, with no significant instrument  cross-normalisation required (Figure~\ref{fig:BATspec}). 
The intrinsic nuclear luminosities for this model are $L_{2-10}=4.1 \times 10^{42} {\rm erg\, s^{-1}}$, 
$L_{10-50}=1.2 \times 10^{43} {\rm erg\, s^{-1}}$.
 
 \subsection{Patchy Torus Model}
The factor of 3 scaling factor for the scattered spectrum  indicates the need for a more complex version of this model. One possibility is that a substantial fraction of the direct emission is hidden from our sight-line by a Compton-thick absorber. In such a case, the scattered component results from a luminosity a factor of $\sim 3$ higher than the observed X-ray continuum. 

We therefore constructed a model that allows the intrinsic continuum to pass through two different sight-lines, of independent column densities, to mimic a more complex case, i.e. a patchy torus reprocessor. The model therefore has two line-of-sight absorbing zones, each with  global scattering gas that is coupled in column density to each of the line-of-sight absorbers. The global/scattering components are inclined at $\theta=0^{\rm o}$, as for some previous models. 

This provided a good fit to the data (Table~3).  A small amount (1.5\%) of continuum is  allowed to leak through unabsorbed to explain the soft excess, as found for previous models. Of the remainder of the continuum, 85\% is viewed through a column density of $\sim 1.5  \times 10^{24}{\rm cm^{-2}}$, 15\% through a much lower column of $\sim 3  \times 10^{23}{\rm cm^{-2}}$. Each of these line-of-sight absorbers is allowed an associated global scattering component with a scaling factor frozen at 1.0. This fit works very well ($\chi^2=526/436$), relieving the need for arbitrary scaling of the global scattered component.

\begin{table*}
	\centering
	\caption{Torus Model Fits }
	\label{tab:table}
	\begin{tabular}{  |   l l l l l l  |} 
\hline
Par & {\sc coupled i}  & {\sc coupled ii}    & {\sc decoupled i}  & {\sc decoupled ii} & Patchy Torus: {\sc coupled iii} \\  
 \hline
  & & & & & \\
 $\Gamma$ & 1.40 ($<1.41$) & 1.40($<1.41$) & $2.06^{+0.05}_{-0.05}$   & $1.87^{+0.06}_{-0.08}$ & $1.42^{+0.06}_{-0.0p}$ \\
  & & & & & \\
 $N_{H, abs}$ & $0.72^{+0.02}_{-0.04}$  & $0.84^{+0.04}_{-0.03}$ & $1.20^{+0.08}_{-0.05}$  &  $0.81^{+0.10}_{-0.12}$  & $1.54^{+0.50}_{-0.25}$ (85\%) \\
  $/10^{24}{\rm cm^{-2}}$ &   &  &   &   & 0.34$^{+0.11}_{-0.10}$  (15\%) \\
 &   & & & & \\
  & & & & & \\
  $N_{H,scatt}$ & 0.72(l) & 0.84(l) & 9.91($>7.63$)  & 9.02 ($>4.18$) & 1.54(l) \\
 $/10^{24}{\rm cm^{-2}}$ & & & & & 0.34(l) \\
 & &  &  & & \\
 & & & & & \\
  & & & & & \\
 $ \theta_{scatt}$ & 90$^{\rm o}$(f) & 70$^{\rm o}$(f) & 0(f)  &   0(f) & 0(f)/0(f) \\
  & & & & & \\
 $C  $ & 1.0(f)  & 1.0(f)   & 1.0(f)  &  $2.61^{+0.65}_{-0.45}$ & 1.0(f)/1.0(f) \\
   & & & & & \\
 $PL_{unabs}$ & $2.91^{+0.25}_{-0.18}$\% &  $2.00^{+0.20}_{-0.19}$ &  $0.21^{+0.05}_{-0.05}$  & $0.92^{+0.51}_{-0.30}$ &  $1.51^{+0.30}_{-0.29}$  \\
  \% & & & &  & \\
  & & & & & \\
 $N_{H,Gal}$ & $2.04^{+0.49}_{-0.44}$ &  $1.72^{+0.47}_{-0.43}$ &  $2.30^{+0.50}_{-0.0p}$  &  $1.95^{+0.49}_{-0.46}$ &  $1.02^{+0.48}_{-0.42}$  \\
 $/10^{21}{\rm cm^{-2}}$ & &  &  &  & \\
  & & & & & \\
 $\chi^2$ & 709/438   & 698/438  & 596/437 &   520/436 &  526/436 \\
  & & & & & \\
\hline
\end{tabular}
\begin{tablenotes}
      \item 	\ \ \ \ \ \ \ \ \ \ \ \ \ \ \ \ \ \ \ \  \ \ \ \ \ \ \ \ \ \  \ \ \ \ \ \ \ \ \ \ \ \ \ \ \ \ \ \ \ \ \ \ \  $p$ indicates the parameter pegged at the preset limit
     \item 	\ \ \ \ \ \ \ \ \ \ \ \ \ \ \ \ \ \ \ \  \ \ \ \ \ \ \ \ \ \  \ \ \ \ \ \ \ \ \ \ \ \ \ \ \ \ \ \ \ \   \ \ \ $f$ indicates the parameter was fixed 
      \item 	\ \ \ \ \ \ \ \ \ \ \ \ \ \ \ \ \ \ \ \  \ \ \ \ \ \ \ \ \ \  \ \ \ \ \ \ \ \ \ \ \ \ \ \ \ \ \ \ \ \  \ \ \  $l$ indicates the parameter was linked (see text for details)
         \end{tablenotes}
        \end{table*}

\begin{figure}
\includegraphics[scale=0.43,width=8cm, height=8cm,angle=-90]{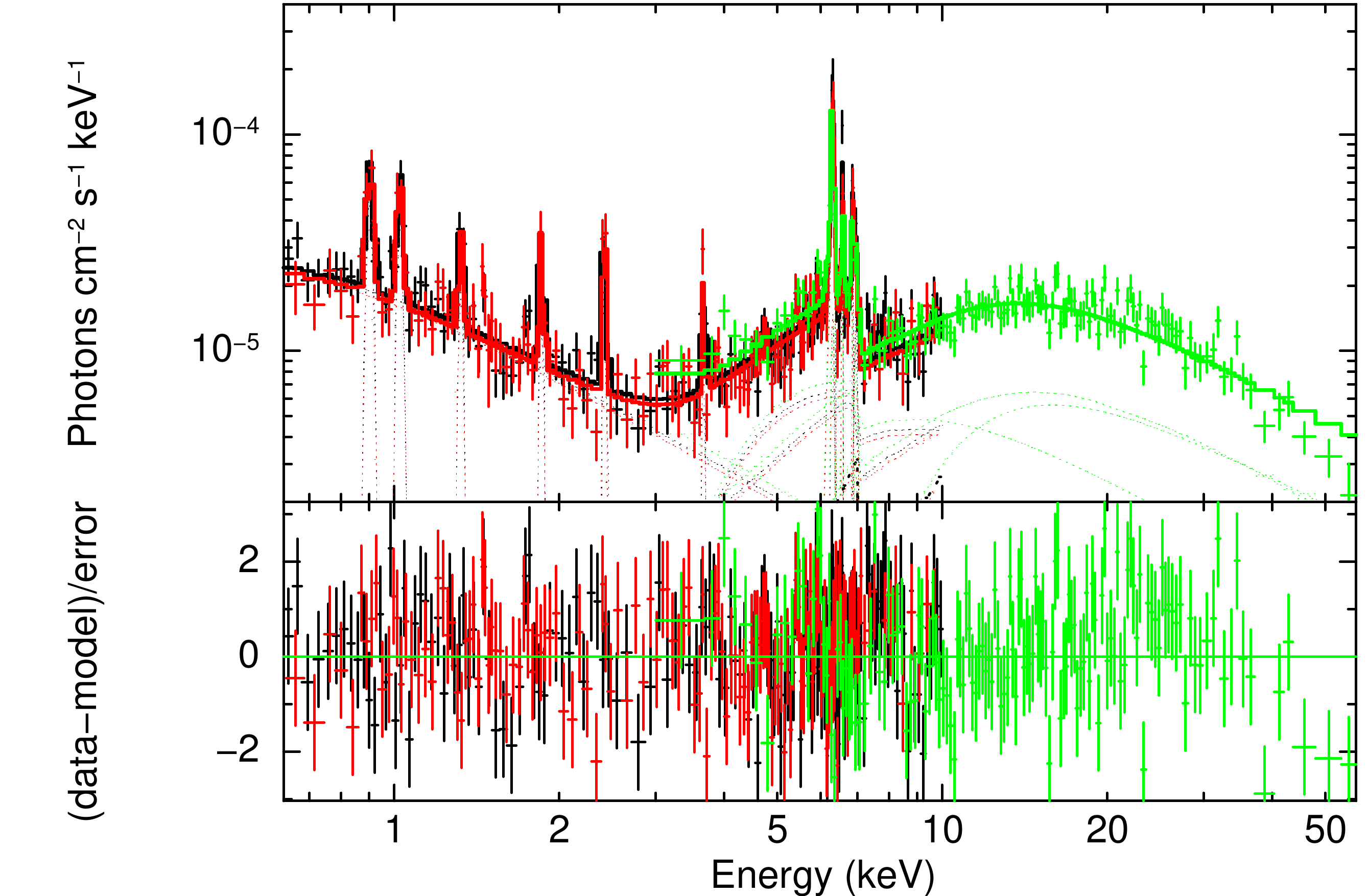}
\caption{ {\it XMM-Newton}/{\it NuSTAR} data fit to the double torus model,  Coupled  {\sc iii}. Colors as for Figure~\ref{fig:broadfit}.}
\label{fig:twocolumn}
\end{figure}

\section{Discussion}

The presence of Compton-thick absorption appears to be  
common in maser AGN \citep{greenhill08a}. The maser source NGC 1194 is no exception, with 
Compton-thick reprocessing  gas revealed in  {\it NuSTAR} \citep{masini16a} and {\it Suzaku} \citep{tanimoto18a} data. 
As also expected for maser AGN, NGC 1194 is highly inclined to our line-of-sight. There is  significant reddening
of the host galaxy continuum and the narrow line region emission lines. The 
 optical line details have led to a Seyfert 1.9 classification, indicating that a small amount of light can be seen 
directly from the optical broad line region  and, in turn, that the nuclear obscuring gas might not be homogeneous.

New data from simultaneous 2020 {\it XMM-Newton} and  {\it NuSTAR} observations of NGC 1194 span the energy range 0.6 - 50 keV.  
The {\it XMM-Newton} spectrum shows many strong soft-band emission lines that 
are a superposition of contributions from ionized extended gas mapped in {\it Chandra} images  (Braito et al., in prep) and neutral lines, 
likely dominated by contributions from a toroidal nuclear reprocessor and the narrow line region gas. 

The  coupled {\sc mytorus} model was applied to  account for the broad X-ray form of the combined 2020 data. The single coupled model configuration does not provide a good fit to the data as the curvature below 10 keV requires a lower line-of-sight column density than that which is required to explain the observed Fe K emission line and overall reflection signature strength from global gas out of the line-of-sight. Further to this, the single zone  coupled models are not found to be a statistically acceptable fit to the data. While the coupled model provided an acceptable fit to the shorter, archived {\it NuSTAR} observation from 2015, we found that the model fit to that  earlier epoch  under-predicted the X-ray flux below 3 keV that was sampled by our {\it XMM-Newton}/{\it NuSTAR} campaign. The availability of simultaneous data down  to lower X-ray energies drives the need for a more complex model than was previously published for this AGN. 

Decoupling the column density for reflection and absorption greatly improves the {\sc mytorus} fit, such that the shape of the model spectrum is well matched  to the data. However, while decoupled {\sc ii} provides a good statistical  fit, the X-ray continuum required to provide the strength of the global reflection observed, is a 
factor of $2.61^{+0.65}_{-0.45}$ higher than the continuum observed (Table~3, column 5, fit parameter C). As NGC 1194 does not show a variable continuum,  this enhanced reflection cannot be attributed to a previous, unobserved epoch of higher continuum.  It may be that this higher continuum is present, but that Compton thick gas is responsible for hiding the full continuum from direct view. 

If, indeed, a large fraction of the continuum is hidden from view, then examination of the spectral energy distribution (SED) for NGC 1194 might help confirm that hypothesis.   In a study of SEDs for a sample of Compton-thick AGN, \citet{brightman17a} calculate the bolometric correction factors 
 $\kappa_{\rm BOL}= log \frac {L_{\rm BOL}}{L_X}$  where $L_{BOL}$ and $L_X$ are  the bolometric and the 2-10 keV luminosity. Taking the intrinsic X-ray luminosity to be 
 $L_{2-10}=4.1 \times 10^{42} {\rm erg\, s^{-1}}$, from the {\sc decoupled ii} fit, and the bolometric luminosity from \citet{brightman17a} to be $L_{BOL}=5.5 \times 10^{44}  {\rm erg\, s^{-1}}$, we estimate 
$\kappa_{\rm BOL}\sim 2.13$. This value is similar to that found by  \citet{brightman17a} and higher than expected compared to the rest of their Compton thick sample. 

 \citet{brightman17a} explore the range of $\kappa_{\rm BOL}$ values exhibited, as a function of 
Eddington ratio ($\lambda_{Edd} $). Based on the sample distribution and a value of $\lambda_{Edd} =0.08 $ for 
NGC 1194 \citep{brightman17a}, the value of $\kappa_{\rm BOL}$ should lie within $0.8 - 1.5$ for NGC 1194. The initial estimation
$\kappa_{\rm BOL}\sim 2.13$ is significantly higher than the expected range. 
If we apply a correction to the X-ray continuum luminosity, based upon the {\sc decoupled ii} fit, then (propagating the uncertainties on that scalar, Table 3, column 5) we estimate a corrected 
$\kappa_{\rm BOL}$ in the range  1.61-1.79, much closer 
to the expected  value for NGC 1194, based on the \citet{brightman17a} sample study.  
 
 An alternative model using two zones of reprocessing gas allows representation of a more complex scenario, such as a patchy torus.
 In the context of such,  85\% of sight-lines to the nucleus are through Compton-thick gas and 15\% through Compton-thin gas. Such a structure effectively gives partial-covering of the  X-ray continuum, which is a common observational phenomenon in local AGN \citep[e.g.][and references therein]{turner11a,risaliti09b}.  This result is consistent with  the torus itself being intrinsically patchy, with a range of column densities (from a patchwork of clouds) viewed over different sight-lines. 
Alternatively the Compton-thin gas could  arise from the BLR region,  located closer in and viewed through holes in a Compton thick torus.

\section{Conclusions}

We have performed  broad band X-ray spectroscopy of a joint  {\it XMM-Newton} and  {\it NuSTAR} observation of the Seyfert 1.9 galaxy, NGC 1194. The source flux is steady and the spectrum shows strong line emission and a very hard  form. The spectrum can be described by torus models with solutions indicating the global column density to be an order of magnitude higher than the line-of-sight absorption. 
In such models, the reflection strength is a factor $\sim 3$ higher than would be nominally expected.  Such a result may indicate that a large fraction of continuum photons are hidden from view in the {\it XMM-Newton}/{\it NuSTAR} bandpass. This idea is supported by the atypical  SED of the source. 

Alternatively, the spectral form can be modeled using two toroidal reprocessor components that could represent a more complex structure, such as a patchy torus. In this case    85\%  of the sight-lines through this patchy torus are viewed through Compton-thick gas and the remaining 15\% through  Compton-thin gas.  

Both of these torus model solutions are based upon an assumption of a solar abundance of Fe. It is of particular interest that the application of  the patchy torus model alleviates the need for a significantly greater intrinsic luminosity to be seen by the reprocessor, than is viewed by the observer. This is not an intuitive result and this insight could not have been obtained by the application of ad hoc models to these data. 

\section*{Acknowledgements}

This paper is dedicated to the memory of Ian Michael George (1963-2020). The best colleague and friend we could have ever hoped for.
\vspace{0.1mm}

\noindent TJT acknowledges support from NASA grant  80NSSC20K0028.   Valentina Braito and Paola Severgnini acknowledge financial contribution from the agreements
ASI-INAF n.2017-14-H.0. We are  grateful to the {\it XMM-Newton} and {\it NuSTAR} operations teams  for performing  this campaign and providing 
software and calibration for the data analysis.

\section*{Data Availability}
This research has  made use of data obtained from the High Energy Astrophysics Science 
Archive Research Center (HEASARC), provided by NASA's Goddard Space Flight Center.




\bibliographystyle{mnras}

\bibliography{xray_Apr2020} 

\begin{thebibliography}{}
\makeatletter
\relax
\def\mn@urlcharsother{\let\do\@makeother \do\$\do\&\do\#\do\^\do\_\do\%\do\~}
\def\mn@doi{\begingroup\mn@urlcharsother \@ifnextchar [ {\mn@doi@}
  {\mn@doi@[]}}
\def\mn@doi@[#1]#2{\def\@tempa{#1}\ifx\@tempa\@empty \href
  {http://dx.doi.org/#2} {doi:#2}\else \href {http://dx.doi.org/#2} {#1}\fi
  \endgroup}
\def\mn@eprint#1#2{\mn@eprint@#1:#2::\@nil}
\def\mn@eprint@arXiv#1{\href {http://arxiv.org/abs/#1} {{\tt arXiv:#1}}}
\def\mn@eprint@dblp#1{\href {http://dblp.uni-trier.de/rec/bibtex/#1.xml}
  {dblp:#1}}
\def\mn@eprint@#1:#2:#3:#4\@nil{\def\@tempa {#1}\def\@tempb {#2}\def\@tempc
  {#3}\ifx \@tempc \@empty \let \@tempc \@tempb \let \@tempb \@tempa \fi \ifx
  \@tempb \@empty \def\@tempb {arXiv}\fi \@ifundefined
  {mn@eprint@\@tempb}{\@tempb:\@tempc}{\expandafter \expandafter \csname
  mn@eprint@\@tempb\endcsname \expandafter{\@tempc}}}

\bibitem[\protect\citeauthoryear{{Anders} \& {Grevesse}}{{Anders} \&
  {Grevesse}}{1989}]{anders89a}
{Anders} E.,  {Grevesse} N.,  1989, \mn@doi [\gca]
  {10.1016/0016-7037(89)90286-X}, \href
  {http://adsabs.harvard.edu/abs/1989GeCoA..53..197A} {53, 197}

\bibitem[\protect\citeauthoryear{{Balokovi{\'c}} et~al.,}{{Balokovi{\'c}}
  et~al.}{2018}]{balokovic18a}
{Balokovi{\'c}} M.,  et~al., 2018, \mn@doi [\apj] {10.3847/1538-4357/aaa7eb},
  \href {http://adsabs.harvard.edu/abs/2018ApJ...854...42B} {854, 42}

\bibitem[\protect\citeauthoryear{{Blustin}, {Page}, {Fuerst},
  {Branduardi-Raymont}  \& {Ashton}}{{Blustin} et~al.}{2005}]{blustin05a}
{Blustin} A.~J.,  {Page} M.~J.,  {Fuerst} S.~V.,  {Branduardi-Raymont} G.,
  {Ashton} C.~E.,  2005, \mn@doi [\aap] {10.1051/0004-6361:20041775}, \href
  {http://adsabs.harvard.edu/abs/2005A%26A...431..111B} {431, 111}

\bibitem[\protect\citeauthoryear{{Braito}, {Reeves}, {Bianchi}, {Nardini}  \&
  {Piconcelli}}{{Braito} et~al.}{2017}]{braito17a}
{Braito} V.,  {Reeves} J.~N.,  {Bianchi} S.,  {Nardini} E.,   {Piconcelli} E.,
  2017, \mn@doi [\aap] {10.1051/0004-6361/201630322}, \href
  {https://ui.adsabs.harvard.edu/abs/2017A&A...600A.135B} {600, A135}

\bibitem[\protect\citeauthoryear{{Brightman} \& {Nandra}}{{Brightman} \&
  {Nandra}}{2011}]{brightman11a}
{Brightman} M.,  {Nandra} K.,  2011, \mn@doi [\mnras]
  {10.1111/j.1365-2966.2011.18612.x}, \href
  {http://adsabs.harvard.edu/abs/2011MNRAS.414.3084B} {414, 3084}

\bibitem[\protect\citeauthoryear{{Brightman} et~al.,}{{Brightman}
  et~al.}{2015}]{brightman15a}
{Brightman} M.,  et~al., 2015, \mn@doi [\apj] {10.1088/0004-637X/805/1/41},
  \href {https://ui.adsabs.harvard.edu/abs/2015ApJ...805...41B} {805, 41}

\bibitem[\protect\citeauthoryear{{Brightman} et~al.,}{{Brightman}
  et~al.}{2017}]{brightman17a}
{Brightman} M.,  et~al., 2017, \mn@doi [\apj] {10.3847/1538-4357/aa75c9}, \href
  {https://ui.adsabs.harvard.edu/abs/2017ApJ...844...10B} {844, 10}

\bibitem[\protect\citeauthoryear{{Della Ceca} et~al.,}{{Della Ceca}
  et~al.}{2008}]{dellaceca08a}
{Della Ceca} R.,  et~al., 2008, \mn@doi [\aap] {10.1051/0004-6361:20079319},
  \href {https://ui.adsabs.harvard.edu/abs/2008A&A...487..119D} {487, 119}

\bibitem[\protect\citeauthoryear{{Elitzur} \& {Shlosman}}{{Elitzur} \&
  {Shlosman}}{2006}]{elitzur06a}
{Elitzur} M.,  {Shlosman} I.,  2006, \mn@doi [\apjl] {10.1086/508158}, \href
  {http://adsabs.harvard.edu/abs/2006ApJ...648L.101E} {648, L101}

\bibitem[\protect\citeauthoryear{{Georgantopoulos} \&
  {Akylas}}{{Georgantopoulos} \& {Akylas}}{2019}]{georgantopoulos19a}
{Georgantopoulos} I.,  {Akylas} A.,  2019, \mn@doi [\aap]
  {10.1051/0004-6361/201833038}, \href
  {https://ui.adsabs.harvard.edu/abs/2019A&A...621A..28G} {621, A28}

\bibitem[\protect\citeauthoryear{{George} \& {Fabian}}{{George} \&
  {Fabian}}{1991}]{george91a}
{George} I.~M.,  {Fabian} A.~C.,  1991, \mnras, \href
  {http://adsabs.harvard.edu/abs/1991MNRAS.249..352G} {249, 352}

\bibitem[\protect\citeauthoryear{{Greenhill}, {Tilak}  \&
  {Madejski}}{{Greenhill} et~al.}{2008}]{greenhill08a}
{Greenhill} L.~J.,  {Tilak} A.,   {Madejski} G.,  2008, \mn@doi [\apjl]
  {10.1086/592782}, \href
  {https://ui.adsabs.harvard.edu/abs/2008ApJ...686L..13G} {686, L13}

\bibitem[\protect\citeauthoryear{{HI4PI Collaboration}}{{HI4PI
  Collaboration}}{2016}]{bekhti16a}
{HI4PI Collaboration} 2016, \mn@doi [\aap] {10.1051/0004-6361/201629178}, \href
  {https://ui.adsabs.harvard.edu/abs/2016A&A...594A.116H} {594, A116}

\bibitem[\protect\citeauthoryear{{Harrison} et~al.,}{{Harrison}
  et~al.}{2013}]{harrison13a}
{Harrison} F.~A.,  et~al., 2013, \mn@doi [\apj] {10.1088/0004-637X/770/2/103},
  \href {http://adsabs.harvard.edu/abs/2013ApJ...770..103H} {770, 103}

\bibitem[\protect\citeauthoryear{{Ikeda}, {Awaki}  \& {Terashima}}{{Ikeda}
  et~al.}{2009}]{ikeda09}
{Ikeda} S.,  {Awaki} H.,   {Terashima} Y.,  2009, \mn@doi [\apj]
  {10.1088/0004-637X/692/1/608}, \href
  {http://adsabs.harvard.edu/abs/2009ApJ...692..608I} {692, 608}

\bibitem[\protect\citeauthoryear{{Kallman} \& {Bautista}}{{Kallman} \&
  {Bautista}}{2001}]{kallman01a}
{Kallman} T.,  {Bautista} M.,  2001, \mn@doi [\apjs] {10.1086/319184}, \href
  {http://adsabs.harvard.edu/abs/2001ApJS..133..221K} {133, 221}

\bibitem[\protect\citeauthoryear{{Kallman}, {Palmeri}, {Bautista}, {Mendoza}
  \& {Krolik}}{{Kallman} et~al.}{2004}]{kallman04a}
{Kallman} T.~R.,  {Palmeri} P.,  {Bautista} M.~A.,  {Mendoza} C.,   {Krolik}
  J.~H.,  2004, \mn@doi [\apjs] {10.1086/424039}, \href
  {http://adsabs.harvard.edu/abs/2004ApJS..155..675K} {155, 675}

\bibitem[\protect\citeauthoryear{{Kaspi} et~al.,}{{Kaspi}
  et~al.}{2002}]{kaspi02a}
{Kaspi} S.,  et~al., 2002, \mn@doi [\apj] {10.1086/341113}, \href
  {http://adsabs.harvard.edu/abs/2002ApJ...574..643K} {574, 643}

\bibitem[\protect\citeauthoryear{{Kuo} et~al.,}{{Kuo} et~al.}{2011}]{kuo11a}
{Kuo} C.~Y.,  et~al., 2011, \mn@doi [\apj] {10.1088/0004-637X/727/1/20}, \href
  {https://ui.adsabs.harvard.edu/abs/2011ApJ...727...20K} {727, 20}

\bibitem[\protect\citeauthoryear{{Liu} \& {Li}}{{Liu} \& {Li}}{2014}]{liu14a}
{Liu} Y.,  {Li} X.,  2014, \mn@doi [\apj] {10.1088/0004-637X/787/1/52}, \href
  {http://adsabs.harvard.edu/abs/2014ApJ...787...52L} {787, 52}

\bibitem[\protect\citeauthoryear{{Madsen}, {Beardmore}, {Forster}, {Guainazzi},
  {Marshall}, {Miller}, {Page}  \& {Stuhlinger}}{{Madsen}
  et~al.}{2017}]{madsen17a}
{Madsen} K.~K.,  {Beardmore} A.~P.,  {Forster} K.,  {Guainazzi} M.,  {Marshall}
  H.~L.,  {Miller} E.~D.,  {Page} K.~L.,   {Stuhlinger} M.,  2017, \mn@doi
  [\aj] {10.3847/1538-3881/153/1/2}, \href
  {https://ui.adsabs.harvard.edu/abs/2017AJ....153....2M} {153, 2}

\bibitem[\protect\citeauthoryear{{Marchese}, {Braito}, {Della Ceca},
  {Caccianiga}  \& {Severgnini}}{{Marchese} et~al.}{2012}]{marchese12a}
{Marchese} E.,  {Braito} V.,  {Della Ceca} R.,  {Caccianiga} A.,   {Severgnini}
  P.,  2012, \mn@doi [\mnras] {10.1111/j.1365-2966.2012.20445.x}, \href
  {http://adsabs.harvard.edu/abs/2012MNRAS.421.1803M} {421, 1803}

\bibitem[\protect\citeauthoryear{{Marchesi} et~al.,}{{Marchesi}
  et~al.}{2019}]{marchesi19a}
{Marchesi} S.,  et~al., 2019, \mn@doi [\apj] {10.3847/1538-4357/aafbeb}, \href
  {https://ui.adsabs.harvard.edu/abs/2019ApJ...872....8M} {872, 8}

\bibitem[\protect\citeauthoryear{{Masini} et~al.,}{{Masini}
  et~al.}{2016}]{masini16a}
{Masini} A.,  et~al., 2016, \mn@doi [\aap] {10.1051/0004-6361/201527689}, \href
  {https://ui.adsabs.harvard.edu/abs/2016A&A...589A..59M} {589, A59}

\bibitem[\protect\citeauthoryear{{Murphy} \& {Yaqoob}}{{Murphy} \&
  {Yaqoob}}{2009}]{murphy09a}
{Murphy} K.~D.,  {Yaqoob} T.,  2009, \mn@doi [\mnras]
  {10.1111/j.1365-2966.2009.15025.x}, \href
  {http://adsabs.harvard.edu/abs/2009MNRAS.397.1549M} {397, 1549}

\bibitem[\protect\citeauthoryear{{Nenkova}, {Sirocky}, {Nikutta}, {Ivezi{\'c}}
  \& {Elitzur}}{{Nenkova} et~al.}{2008}]{nenkova08b}
{Nenkova} M.,  {Sirocky} M.~M.,  {Nikutta} R.,  {Ivezi{\'c}} {\v Z}.,
  {Elitzur} M.,  2008, \mn@doi [\apj] {10.1086/590483}, \href
  {http://adsabs.harvard.edu/abs/2008ApJ...685..160N} {685, 160}

\bibitem[\protect\citeauthoryear{{Osterbrock} \& {Ferland}}{{Osterbrock} \&
  {Ferland}}{2006}]{osterbrock06a}
{Osterbrock} D.~E.,  {Ferland} G.~J.,  2006, {Astrophysics of gaseous nebulae
  and active galactic nuclei}.
Astrophysics of gaseous nebulae and active galactic nuclei, 2nd.~ed.~by
  D.E.~Osterbrock and G.J.~Ferland.~Sausalito, CA: University Science Books,
  2006

\bibitem[\protect\citeauthoryear{{Reeves}, {Nandra}, {George}, {Pounds},
  {Turner}  \& {Yaqoob}}{{Reeves} et~al.}{2004}]{reeves04a}
{Reeves} J.~N.,  {Nandra} K.,  {George} I.~M.,  {Pounds} K.~A.,  {Turner}
  T.~J.,   {Yaqoob} T.,  2004, \mn@doi [\apj] {10.1086/381091}, \href
  {http://adsabs.harvard.edu/abs/2004ApJ...602..648R} {602, 648}

\bibitem[\protect\citeauthoryear{{Risaliti} et~al.,}{{Risaliti}
  et~al.}{2009}]{risaliti09b}
{Risaliti} G.,  et~al., 2009, \mn@doi [\mnras]
  {10.1111/j.1745-3933.2008.00580.x}, \href
  {http://adsabs.harvard.edu/abs/2009MNRAS.393L...1R} {393, L1}

\bibitem[\protect\citeauthoryear{{Savage} \& {Mathis}}{{Savage} \&
  {Mathis}}{1979}]{savage79a}
{Savage} B.~D.,  {Mathis} J.~S.,  1979, \mn@doi [\araa]
  {10.1146/annurev.aa.17.090179.000445}, \href
  {https://ui.adsabs.harvard.edu/abs/1979ARA&A..17...73S} {17, 73}

\bibitem[\protect\citeauthoryear{{Severgnini}, {Caccianiga}  \& {Della
  Ceca}}{{Severgnini} et~al.}{2012}]{severgnini12a}
{Severgnini} P.,  {Caccianiga} A.,   {Della Ceca} R.,  2012, \mn@doi [\aap]
  {10.1051/0004-6361/201118417}, \href
  {https://ui.adsabs.harvard.edu/abs/2012A&A...542A..46S} {542, A46}

\bibitem[\protect\citeauthoryear{{Shull} \& {van Steenberg}}{{Shull} \& {van
  Steenberg}}{1985}]{shull85a}
{Shull} J.~M.,  {van Steenberg} M.~E.,  1985, \mn@doi [\apj] {10.1086/163327},
  \href {https://ui.adsabs.harvard.edu/abs/1985ApJ...294..599S} {294, 599}

\bibitem[\protect\citeauthoryear{{Tanimoto}, {Ueda}, {Kawamuro}, {Ricci},
  {Awaki}  \& {Terashima}}{{Tanimoto} et~al.}{2018}]{tanimoto18a}
{Tanimoto} A.,  {Ueda} Y.,  {Kawamuro} T.,  {Ricci} C.,  {Awaki} H.,
  {Terashima} Y.,  2018, \mn@doi [\apj] {10.3847/1538-4357/aaa47c}, \href
  {https://ui.adsabs.harvard.edu/abs/2018ApJ...853..146T} {853, 146}

\bibitem[\protect\citeauthoryear{{Tatum}, {Turner}, {Miller}  \&
  {Reeves}}{{Tatum} et~al.}{2013}]{tatum13a}
{Tatum} M.~M.,  {Turner} T.~J.,  {Miller} L.,   {Reeves} J.~N.,  2013, \mn@doi
  [\apj] {10.1088/0004-637X/762/2/80}, \href
  {http://adsabs.harvard.edu/abs/2013ApJ...762...80T} {762, 80}

\bibitem[\protect\citeauthoryear{{Turner}, {Miller}, {Kraemer}  \&
  {Reeves}}{{Turner} et~al.}{2011}]{turner11a}
{Turner} T.~J.,  {Miller} L.,  {Kraemer} S.~B.,   {Reeves} J.~N.,  2011,
  \mn@doi [\apj] {10.1088/0004-637X/733/1/48}, \href
  {http://adsabs.harvard.edu/abs/2011ApJ...733...48T} {733, 48}

\bibitem[\protect\citeauthoryear{{Verner}, {Ferland}, {Korista}  \&
  {Yakovlev}}{{Verner} et~al.}{1996}]{verner96a}
{Verner} D.~A.,  {Ferland} G.~J.,  {Korista} K.~T.,   {Yakovlev} D.~G.,  1996,
  \mn@doi [\apj] {10.1086/177435}, \href
  {http://adsabs.harvard.edu/abs/1996ApJ...465..487V} {465, 487}

\bibitem[\protect\citeauthoryear{{V{\'e}ron-Cetty} \&
  {V{\'e}ron}}{{V{\'e}ron-Cetty} \& {V{\'e}ron}}{2006}]{veroncetty06a}
{V{\'e}ron-Cetty} M.~P.,  {V{\'e}ron} P.,  2006, \mn@doi [\aap]
  {10.1051/0004-6361:20065177}, \href
  {https://ui.adsabs.harvard.edu/abs/2006A&A...455..773V} {455, 773}

\bibitem[\protect\citeauthoryear{{Wilms}, {Allen}  \& {McCray}}{{Wilms}
  et~al.}{2000}]{wilms2000a}
{Wilms} J.,  {Allen} A.,   {McCray} R.,  2000, \mn@doi [\apj] {10.1086/317016},
  \href {https://ui.adsabs.harvard.edu/abs/2000ApJ...542..914W} {542, 914}

\bibitem[\protect\citeauthoryear{{Yaqoob}}{{Yaqoob}}{2012}]{yaqoob12a}
{Yaqoob} T.,  2012, \mn@doi [\mnras] {10.1111/j.1365-2966.2012.21129.x}, \href
  {https://ui.adsabs.harvard.edu/abs/2012MNRAS.423.3360Y} {423, 3360}

\bibitem[\protect\citeauthoryear{{Yaqoob}, {Murphy}, {Miller}  \&
  {Turner}}{{Yaqoob} et~al.}{2010}]{yaqoob10a}
{Yaqoob} T.,  {Murphy} K.~D.,  {Miller} L.,   {Turner} T.~J.,  2010, \mn@doi
  [\mnras] {10.1111/j.1365-2966.2009.15657.x}, \href
  {http://adsabs.harvard.edu/abs/2010MNRAS.401..411Y} {401, 411}

\bibitem[\protect\citeauthoryear{{Yaqoob}, {Tatum}, {Scholtes}, {Gottlieb}  \&
  {Turner}}{{Yaqoob} et~al.}{2015}]{yaqoob15a}
{Yaqoob} T.,  {Tatum} M.~M.,  {Scholtes} A.,  {Gottlieb} A.,   {Turner} T.~J.,
  2015, \mn@doi [\mnras] {10.1093/mnras/stv2021}, \href
  {http://adsabs.harvard.edu/abs/2015MNRAS.454..973Y} {454, 973}

\makeatother
\end{thebibliography}





\bsp	
\label{lastpage}
\end{document}